\documentclass[onecolumn,a4paper,12pt,oneside]{article} 
\usepackage[top=2.5cm,left=2.5cm,right=2.5cm,bottom=2.5cm]{geometry}    
\usepackage[nostamp]{draftwatermark} % watermark programmed but not printed
\SetWatermarkText{\bf Preprint}
\SetWatermarkAngle{45}
\SetWatermarkScale{4}
\SetWatermarkLightness{0.85}     
\usepackage{graphicx}
\usepackage{amssymb}
\usepackage{amsmath}
\usepackage{amsfonts}
\usepackage{amsthm}
\usepackage{epstopdf}
\usepackage[usenames,dvipsnames,svgnames,table]{xcolor}
\usepackage{tikz}
\usepackage{array}
\usepackage[T1]{fontenc}
\usepackage[utf8]{inputenc}
\usepackage{booktabs} % for much better looking tables
\usepackage{array} % for better arrays (eg matrices) in maths
\usepackage{paralist} % very flexible & customisable lists (eg. enumerate/itemize, etc.)
\usepackage{listings} 
\usepackage{verbatim} % adds environment for commenting out blocks of text & for better verbatim
\usepackage[colorlinks=true]{hyperref}
\hypersetup{colorlinks=true, linktocpage=true, linkcolor=Blue, citecolor=Green,
hypertexnames=true, urlcolor=MidnightBlue,pdftex}
\usepackage[english]{babel}
\usepackage{wrapfig}
\usepackage{multirow}
\usepackage{quoting}
\usepackage{rotating}
\quotingsetup{font=normalsize}
\theoremstyle{plain}

\usepackage{bm}
\usepackage{abstract}
\usepackage[authoryear]{natbib}
\usepackage{float}
\usepackage{textcomp} 	
\usepackage{tabularx} % Better tables
\usepackage{caption}
\usepackage{authblk} % for authors, affiliations etc.
\usepackage{eso-pic} % for background figures
\usepackage{enumitem} % betters lists
\usepackage{lineno} % for numbering lines
\providecommand{\keywords}[1]{\textbf{{Key words: }} #1} % command for keywords
\usepackage{subfigure}
\title{\large{\bf Experimental study of the wind pressure field on the Notre Dame Cathedral in Paris\\ \bigskip {\it PREPRINT}}}

\author[1]{\footnotesize{Claudio Mannini}}
\author[1]{\footnotesize{Tommaso Massai}}
\author[2]{\footnotesize{Enrico Panettieri}}
\author[1]{\footnotesize{Niccolò Barni}}
\author[1]{\footnotesize{Andrea Giachetti}}
\author[3]{\footnotesize{Margherita Ferrucci}}
\author[2]{\footnotesize{Marco Montemurro}}
\author[4]{\footnotesize{P. Vannucci}\footnote{Corresponding author:  \href{mailto:paolo.vannucci@uvsq.fr}{paolo.vannucci@uvsq.fr}}}

\affil[1]{\footnotesize{CRIACIV-Department of Civil and Environmental Engineering, University of Florence,

Via S. Marta, 3, 50139 Florence, Italy}}
\affil[2]{\footnotesize{Arts et Métiers Institut of Technology - Université de Bordeaux, CNRS, INRA, Bordeaux INP, HESAM Université, I2M UMR5295,  33405 Talence, France.}}
\affil[3]{\footnotesize{ Laboratorio di Fisica Tecnica Ambientale, Università IUAV, 30135 Venezia, Italy.}}
\affil[4]{\footnotesize{LMV - UMR8100, Université Paris-Saclay - UVSQ. 45, Avenue des Etats-Unis,

 78035 Versailles, France.}}

\begin{document}
\maketitle

%%%%%%%%%%%% ABSTRACT %%%%%%%%%%%%%%%%%
\hrule
\begin{abstract}
The paper concerns an experimental study on the wind pressures over the surface of a worldwide known Gothic Cathedral: Notre Dame of Paris. The experimental tests have been conducted in the CRIACIV wind tunnel, Prato (Italy), on a model of the Cathedral at the scale 1:200 reproducing the atmospheric boundary layer. Two types of tests have been conducted: with or without the surrounding modeling the part of the city of Paris near the Cathedral. This has been done, on the one hand, for evaluating the effect of the surrounding buildings onto the wind pressure distribution on the Cathedral, and, on the other hand, to have a wind pressure distribution plausible for any other Cathedral with a similar shape. The tests have been done for all the wind directions and the mean and peak pressures have been recorded.
The results emphasize that the complex geometry of this type of structures is responsible for a peculiar aerodynamic behavior that does not allow estimating correctly the wind loads on the various parts of the Cathedral based on codes and standards, which are tailored for ordinary regular buildings. % 183 parole

% keywords
\end{abstract}
\medskip
\hrule
\bigskip

\bibpunct{(}{)}{,}{a}{}{;}
%%%%%%%%%%% TEXT of The ARTICLE %%%%%%%%%%%%

\section{Introduction}
\label{sec:intro}

The scientific research on  built heritage is more and more  pushed by  environmental problems, such as the climate changes, \citet{Scott_2021}, which represents an increasing threat for its preservation. 
Some recent, still ongoing, research projects testify of the importance of the relationship between built heritage and climate change and of the increasing interest of  governments, supranational organizations, conscious that the conservation of the architectural heritage is linked to the fundamental values of the social and intellectual life of the nations,  \citet{Taher2020}.
We recall, for instance, the projects funded by the European Union within the Joint Programming Initiative on Cultural Heritage and Global Change and by the Cultural Heritage H2020-EU\footnote{HYPERION (Development of a Decision Support System for Improved Resilience and Sustainable Reconstruction of historic areas to cope with Climate Change and Extreme Events based on Novel Sensors and Modelling Tools);\\ ARCH (Advancing Resilience of Historic Areas against Climate-related and other Hazards); \\ CONSECH20 (CONSErvation of 20th century concrete Cultural Heritage in urban changing environments.}.
In particular, the global warming of Earth is causing more and more frequent and strong wind storms, which constitute today (and will constitute) a severe threat for the conservation of some iconic monumental structures. According to \citet{STEENBERGEN2012178}, the climate change implies an increase up to 2.3\% in the hourly mean wind speed with a return period of 50 years.  
An adequate evaluation of the wind pressure on monumental structures is hence of paramount importance for a correct safety evaluation of such structures in the next future. 

This is even more stringent for high-rise buildings, like the Gothic Cathedrals, which are more exposed to the action of wind, whose speed increases with the height above the ground.
An example of the increasing danger such monuments are exposed to, is the damage to the great rose of the Cathedral of Soissons (France), destroyed by the storm Egon on January 13th, 2017. Some minors damages were also suffered by other great Gothic Cathedrals in France during the severe wind storms of the last days of December 1999, like the Notre Dame Cathedral in Paris, where a wind as strong as 169 km/h was recorded inside the city.
Famous is also the collapse of the Spire of Saint Bonifatius Church in Leeuwarden, The Netherlands, during a wind storm in 1976 .
To this end, this paper focuses on the evaluation of wind pressures on Gothic Cathedrals, which are a major example of the built heritage in Europe. 

Four approaches are used to investigate the wind effects on structures in the atmospheric boundary layer: theoretical studies, full-scale measurements, Computation Fluid Dynamics (CFD) numerical simulations and physical experiments on reduced-scale models. 
Theoretical models concern objects of simple, aerodynamic shape, and they are not suited for complex structures like Gothic Cathedrals.
Full-scale measurements are very expensive and require the possibility of equipping with sensors the considered structure. For these reasons, in wind engineering experimental campaigns {\it in situ} are still scarce and often they serve to corroborate either the results of experiments on reduced-scale models or the results of a computation.   
CFD numerical simulations are more and more used in the study of the wind pressure and velocity fields around buildings. Although they can provide detailed information on the relevant flow variables, their accuracy and reliability are still of concern, especially when the simulations regard buildings with a complicated shape, located in an urban environment, as in the case of Gothic Cathedrals. In such situations, the validation of the studies by full-scale measurements or reduced-scale experimental tests is necessary, \citet{BLOCKEN201469}. 
Therefore, experiments on scale physical models in wind tunnels, used since the early twentieth century, cf. \citet{Ferr_Peter}, remain still today an indispensable and reliable tool for a correct analysis of the wind field around buildings of complicated geometry in an urban landscape. 

Due to their dimensions and geometry, the fine detail and richness of their decorations, their particular structural organization, composed by high vaults, flying buttresses, timber roofing, and also due to the materials they are composed of (i.e., stone, mortar, stain-glasses, wood), Gothic Cathedrals are very peculiar and quite delicate structures. The use of some technical norms, e.g. Eurocode~1, \citet{Eurocode1}, in evaluating  the wind load on them, seems inadequate. Indeed, such norms are conceived for the design of ordinary new buildings, while Gothic Cathedrals are existing and really extra-ordinary constructions. In particular, the wind load prescribed by Eurocode 1 is very simplified and, for a complicated structure, it is not possible to 
know whether it over- or under-estimates the real loading. 
Therefore, it would be pretentious and rather unrealistic to assess the structural safety of a Gothic Cathedral, in respect of the increasingly severe wind storms that are caused by climate changes, using so a rough model. Indeed, a deep scientific investigation of the wind-induced pressure distribution on the surface of a Gothic Cathedral is  necessary for a correct risk assessment of such structures or of some parts of them. This is a scientific challenge that still waits for an adequate response. 

 %\begin{comment}
 \begin{figure}[h]
 \begin{center}
 \includegraphics[width=.8\textwidth]{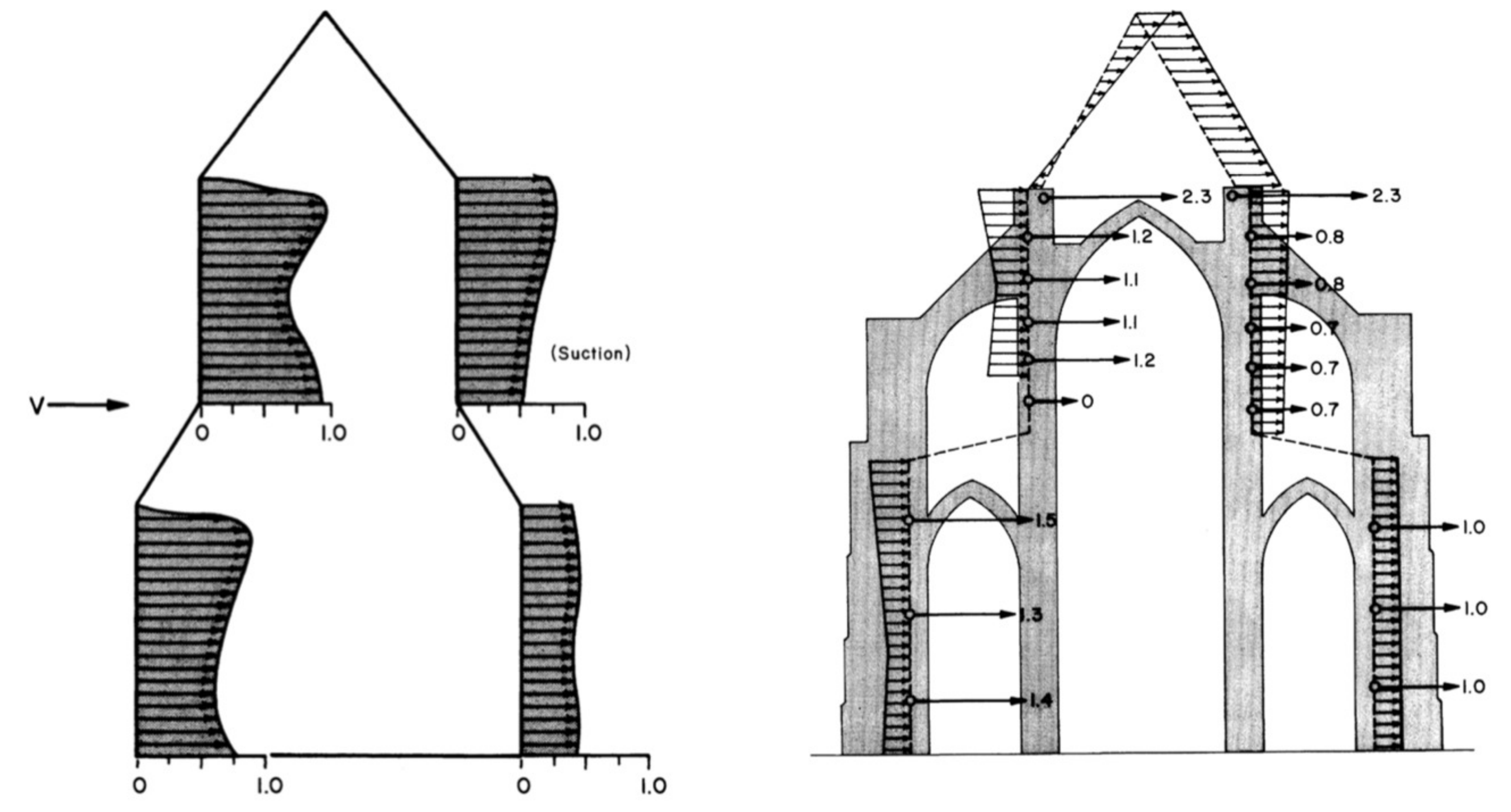}
 \caption{The wind pressure coefficients as experimentally measured in \citet{chien51}, left, and how they are applied to a bi-dimensional Cathedral model in \citet{mark70}, right.}
 \label{fig:1}
 \end{center}
 \end{figure}
%\end{comment}

In the literature, experimental studies of the wind pressure field on a Cathedral, even if not Gothic, are rare. A few works focus on the spires of Cathedrals, or on slender structures similar to spires, or on masonry structures of particular value or architectural interest; some of them are briefly recalled hereafter. 
To the authors' best knowledge, there is only an old wind tunnel investigation on a bi-dimensional model of a Cathedral-like building, \citet{chien51}. Originally, this study did not concern Cathedrals or churches in particular, but just buildings with some typical cross-section shapes, one of them being similar to a simple model of the nave of a Middle Age church. Results were given in the form of bi-dimensional diagrams of the wind pressure coefficient, as shown in Fig.~\ref{fig:1}, and it has been the basis for some subsequent pioneer works on the matter made by R. Marks and co-workers, e.g. \citet{mark70}.  
Two studies on the dynamic response of two slender masonry structures to wind actions, the San Gaudenzio Basilica in Novara and the Mole Antonelliana in Turin (both in Italy), are cited in \citet{CALDERINI2015302}. For the latter, a storm caused the collapse of the spire in 1953, and its reconstruction was undertaken only after wind tunnel tests in the Aeronautics Laboratory of the Turin Polytechnic University in 1954.
The wind pressure on the Church of Sainte-Jeanne-d'Arc in Rouen (France) was determined by wind tunnel experiments in the Eiffel Laboratory in Paris, \citet{LAE_EgliseRouen1972}. 
In \citet{SZALAY1983187},  a wind tunnel study for a damaged ancient church spire is proposed, in order to evaluate the wind load to be used for the design of the reinforcing structure. The experiments were performed in the Hungarian Institute of Building Science.
The investigation about the wind effects on the leaning Tower of Pisa (Italy) is described in \citet{SOLARI1998}. The experiments were devoted to understand the wind load that could cause the collapse of the tower due to the weak strength of the foundation soil. The wind-induced pressures on the monumental roof structure of the XII century Palazzo della Ragione in Padova were also experimentally studied in \citet{BORRI1999}. The tests were performed in the CRIACIV wind tunnel Laboratory in Prato (Italy).
More recently, in a study on the vulnerability of the Capetian architecture to wind storms, \citet{AVACAT2016}, CFD simulations were carried out on 2D simplified models of some churches. In \citet{DOMEDE2019},  the wind load provided by Eurocode~1 is compared to the ancient methods available at the time of the construction (end of XIX century) of the Ile Vierge Lighthouse (France), the tallest stone lighthouse in Europe. 

As far as Gothic Cathedrals are specifically concerned, there are very few studies on the wind actions: this is rather surprising, being these structures sensitive to wind storms due to their height and large dimensions. Normally, the structure of the Cathedral is modeled through a simple bi-dimensional scheme and analyzed with different techniques: the line of pressure method in \citet{ungewitter}; photo-elasticity in \citet{mark70,mark82book,mark84book}; limit analysis in \citet{comobook} and \citet{como15a}. The only work on the wind strength of a Gothic Cathedral considering a three-dimensional, though partial, model and a nonlinear material behavior is a recent paper, \citet{vannucci19}. Just as an example, Fig.~\ref{fig:1} shows the wind load on a Cathedral-like construction as suggested in \citet{chien51} and how this is applied to a bi-dimensional Cathedral model in \citet{mark70}.
In all of the works cited above, the wind pressure is simply modeled as a uniform or linearly variable load. Also in a recent study about the roofing structure of Notre Dame in Paris, destroyed by the fire of April 15th, 2019, the wind load is assumed as step-wise uniform, \citet{vannucci21}.

The present work is an experimental study that aims at providing an extensive evaluation of the wind load that can be expected on the various parts of a Gothic Cathedral. Such a study is clearly still missing in the literature and without any doubt it is a fundamental step for a correct evaluation of the structural safety of such monuments. The experimental investigation is carried out in a boundary-layer wind tunnel on a physical scale model of the Notre Dame Cathedral in Paris, an emblematic building of the French Gothic age. 
%The details of the model, of the experimental tests and of the results are given below.

%\section{Research aim}

The goal of the present study is twofold. Firstly, it aims at providing a precise portrait of the wind pressure field on this iconic construction, which has been the objective of many discussions and structural analyses after the infamous recent fire that destroyed the roof, the spire and damaged parts of the vaults. 

Secondly, regarding the assessment of the wind effects, Notre Dame in Paris can be seen as a sort of paradigm of a more generic Gothic Cathedral. Thus, an emphasis is put on the influence  of the specific surrounding built environment on the pressure field and, by consequence, the tests are made with or without the surrounding city environment. 

It is worth noting that this experimental study is performed not only on an extremely precise physical model of the Cathedral, realized through 3D-printing, but also, and this constitutes a scientific primacy, on a highly instrumented model. Thanks to a careful and very dense distribution of pressure captors, a fine reconstruction of the wind load on the various parts of a Gothic Cathedral has thus been possible for the first time.

\section{Fabrication technology and materials of the Cathedral's model}
\label{sec:model_manuf}

The campaign of experimental tests has been done on a physical model of the Cathedral at the scale of 1:200. This scale has been chosen accounting for the dimensions of the Cathedral (130 m long, 45 m wide, 44 m high at the roof's top and 96 m at the spire's top) in order to satisfy a number of requirements.
On the one hand, the model should be as large as possible to allow for a finer reproduction of the complex geometry of the structure and to facilitate the installation of a large number of pressure captors in all of its parts (see Sect. \ref{subsec: 3D model preparation}).
On the other hand, the model must be \textquotedblleft small\textquotedblright \,compared to the wind tunnel test chamber (2.4 m wide and 1.6 m high in the present case) not to alter the flow field around the construction (blockage effect). The generally accepted safe rule is that the area of the blocking obstacle projected in a plane perpendicular to the flow is less than 5\% of the wind tunnel test section.
However, the most cogent physical limitation is often represented by the scale at which it is possible to reproduce in the wind tunnel the target wind flow characteristics (see Sect. \ref{sec:flow}), which must be scaled in the same way as the model of the Cathedral.
Finally, the reduced-scale model is 65 cm long, 22.5 cm wide, 22 cm high (top of the roof), and the spire is 48 cm high.

\subsection{Fabrication technology}

The realization of the physical, reduced-scale model of Notre Dame was conditioned by some technological requirements. First, in order to have experimental results really representing reality, the model had to reproduce the geometry of the Cathedral with a high fidelity. This is quite a hard task for a building like a gothic Cathedral, which has a complex form and is reach of details, like pinnacles, sculptures, etc. Then, the model had to be instrumented with a huge number of wind pressure captors, cf. Sect. \ref{subsec: 3D model preparation}, placed everywhere over the surface of the model. For these reasons, the model was fabricated using the 3D-printing technology, that enables manufacturing possibilities which cannot be achieved through classical processes, cf., e.g., \citet{NGO2018172, WANG2017442, MITCHELL2018606, CANOVICENT2021102378}. 
In particular, it is possible to realize objects not only with very complex shapes, but also composed of different materials. 
Due to the dimensions and the geometrical complexity of the Notre Dame Cathedral model, the FDM (Fused Deposition Modeling) technology was chosen, except for those regions of the Cathedral characterized by very fine details, e.g., the spire, manufactured via the SLA printing technology (stereolithography) and assembled afterwards.

%\begin{comment}
\begin{figure}[h!]
	\begin{center}
		\includegraphics[width=\textwidth]{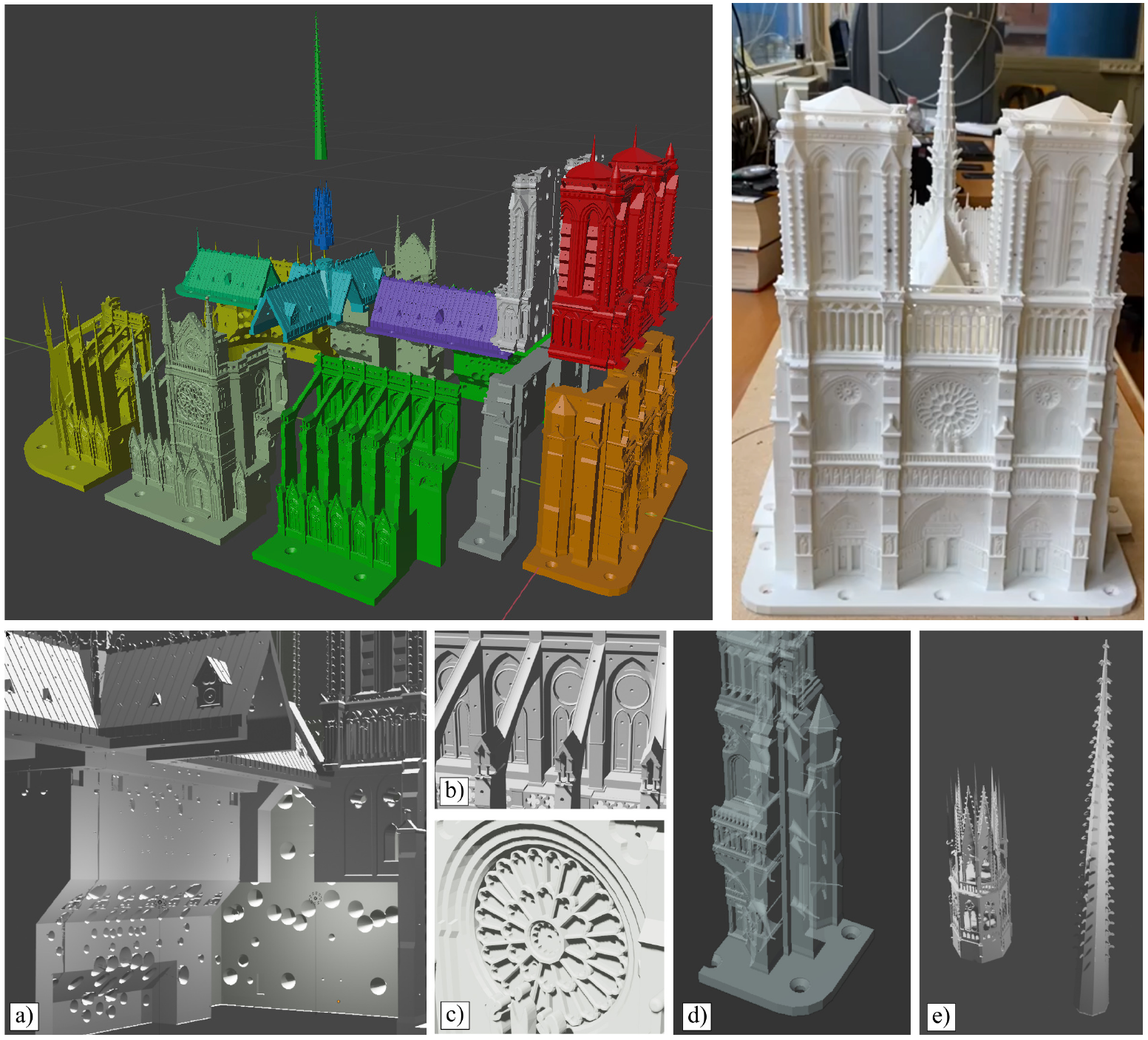}
		\caption{The Cathedral model: top, right: the 15 parts of the numerical model; left: the fabricated model. Bottom: a) detail of the internal volume (façade and nave) with the conical holes to place the pressure tubes; b) detail of the external surface of the nave; c) detail of the rose on the facade; d) internal view of a portion of the facade; e) spire model split into two parts. }
		\label{fig:2}
	\end{center}
\end{figure}

\begin{figure}[h!]
	\begin{center}
		\includegraphics[width=\textwidth]{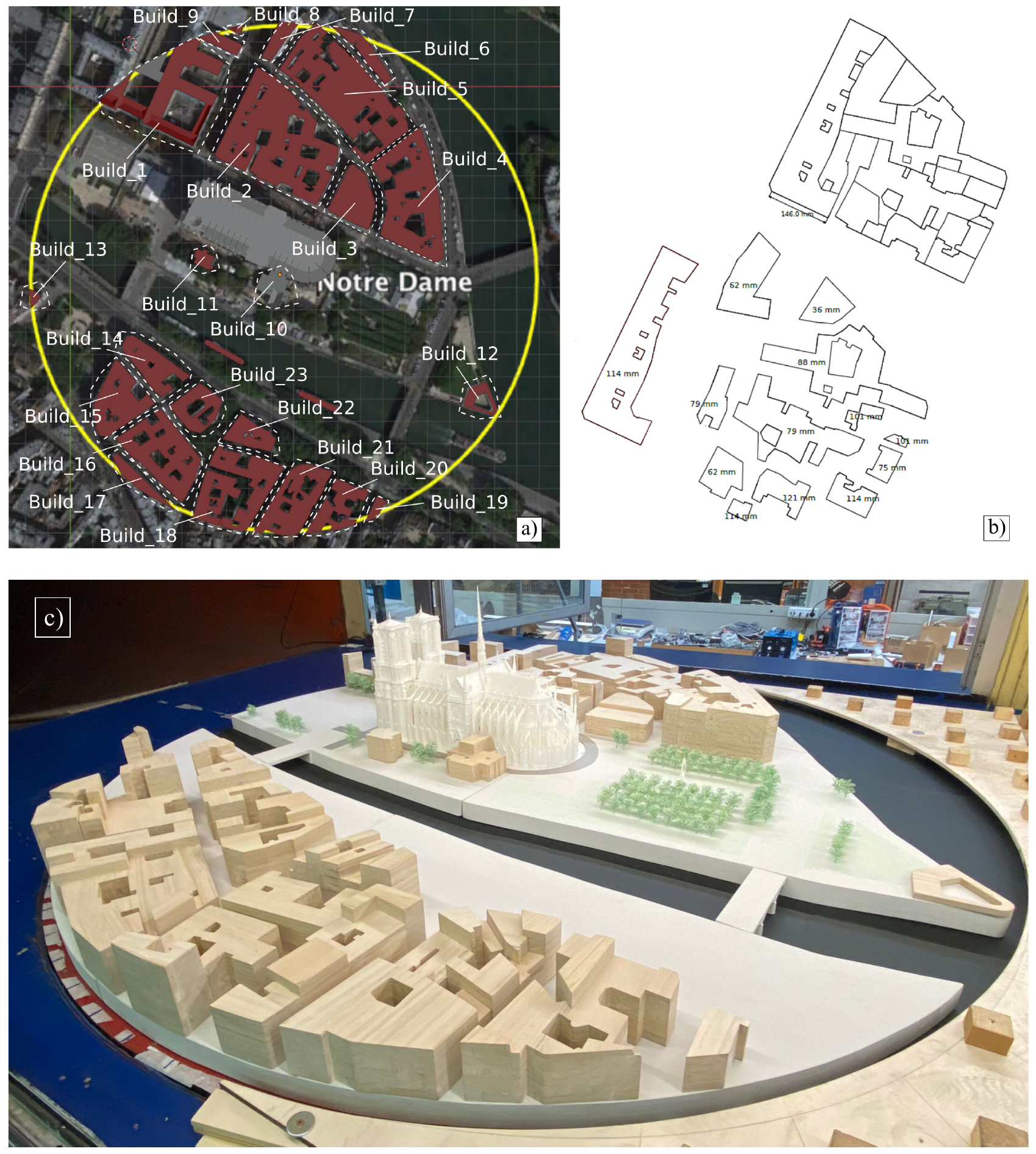}
		\caption{Model of the surrounding of the Cathedral; a) selected blocks of buildings for the model; b) example of a planform with the internal building partitions; c) view of surrounding and Cathedral models mounted on the turning table in the wind tunnel.}
		\label{fig:3}
	\end{center}
\end{figure}
%\end{comment}

\subsection{Model preparation, fabrication and materials}
\label{subsec: 3D model preparation}

The model of the Cathedral has been fabricated starting from an existing high-fidelity numerical mock-up, \citet{NotreDame_model}.
However, the numerical model has been thoroughly modified in order to obtain a physical model of the Cathedral at the scale of 1:200 well adapted to the wind tunnel tests and to comply with the specificity of the FDM printing technology. In fact, the physical model had to be equipped with many pressure captors and it had to allow the different manipulations needed in the laboratory for the set up of the experiment. Also, some modifications have been done on the original numerical model in order to minimize the volume of material and the printing time. 
The 3D computer graphics software Blender,  \citet{BLENDER}, has been used to work on the numerical model of the Cathedral. 

The modifications have been carried out directly on the STL (Standard Tessellation Language) file by means of boolean operations or via adjustments of the initial mesh.
In order to be able to place all of the 1200 flexible tubes that connect the holes on the model surface to the pressure sensors, the walls of the model have been modified on the internal side through the subtraction of a cone, so as to be possible to place the tubes in the holes. This rather complicated operation has been made on the numerical model using a python script coupled with Blender, modifying the file by some boolean operations. 

The spire, which is rich in extremely fine details, has been fabricated using a more precise numerical model, \citet{NotreDame_model_SPIRE}, and a 3D printer based upon the SLA technology, that allows to obtain extremely precise objects, also when of very complicated shape. The model of the spire so obtained has then been assembled with the Cathedral. 

The whole model, Cathedral plus spire, is composed of 15 parts, separately fabricated and then assembled together. This has been done for two reasons: the limits of the printers (maximum height: 600 mm; maximum width: 390 mm) and the need of working on the model for installing the many pressure captors. The different parts have been assembled using magnets and bolts. A scheme of the parts of the  model, some details of it, showing also the holes for the captors, and the final physical model are shown in Fig.~\ref{fig:2}.

The 3D printing of the Cathedral mock-up was performed at the ENSAM facilities in Bordeaux by means of a Lynxter S600D 3D printer, \citet{lynxter}, equipped with a single extrusion filament tool-head, selected in the light of its large building volume and its excellent printing performance.

The 3D printing made use of white PolyMax\texttrademark PLA filament that, thanks to its nano-reinforcement technology, represents a suitable compromise between ease and quality of printing and acceptable mechanical stiffness. The Simplify3D slicing software, \citet{Simplify3D}, was used to prepare the G-code files used by the printing machine. The total printing time was about two weeks.

\subsection{The surrounding environment}

The mock-up of the buildings constituting the surrounding of the Cathedral has been obtained by combining data from two sources: an existing STL file of the center of Paris, cf. the website \citet{3Dcad}, and 3D data extracted from the OpenStreetMap database, \citet{OpenStreetMap}.

For the purposes of the wind tunnel tests, it was sufficient to realize buildings having simplified shapes but correct (scaled) heights. The blocks of buildings constituting the surrounding model are shown in Fig.~\ref{fig:3}(a).

Because of the significant total volume of the buildings and their simple shapes, the model of the surrounding has been fabricated using wood plates, through a milling machine used to cut plates complying with the planforms of the buildings. An example of a planform of an array of buildings is shown in Fig.~\ref{fig:3}(b). For a generic building, the final height was obtained by simply stacking up and gluing together the milled wood plates. 

Finally, the wind tunnel floor was lifted up of 4~cm, so to simulate the presence of the Seine River, being the distance between the base of the Cathedral and the average water level about 8~m at full scale. 
A view of the complete test set-up, with the models of the Cathedral and the surrounding on the turning table of the wind tunnel is shown in Fig.~\ref{fig:3}(c).

\section{Experimental campaign}
\label{sec:exp_camp}

\subsection{General outline}

\begin{figure}[t]
	\begin{center}
	\includegraphics[width=\textwidth]{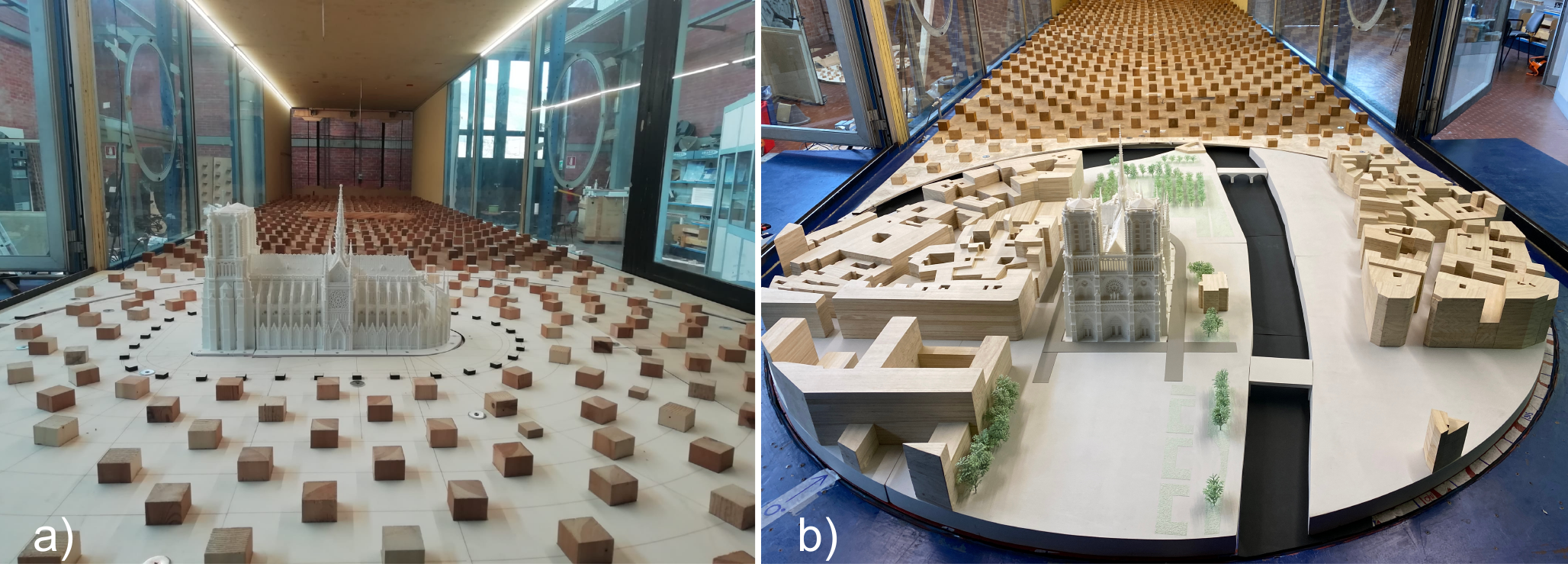}
	\caption{Model of the Cathedral in the wind tunnel: a): without and, b): with surrounding.}
	\label{fig:4}
	\end{center}
\end{figure}

\begin{figure}[h]
	\begin{center}
	\includegraphics[width=0.8\textwidth]{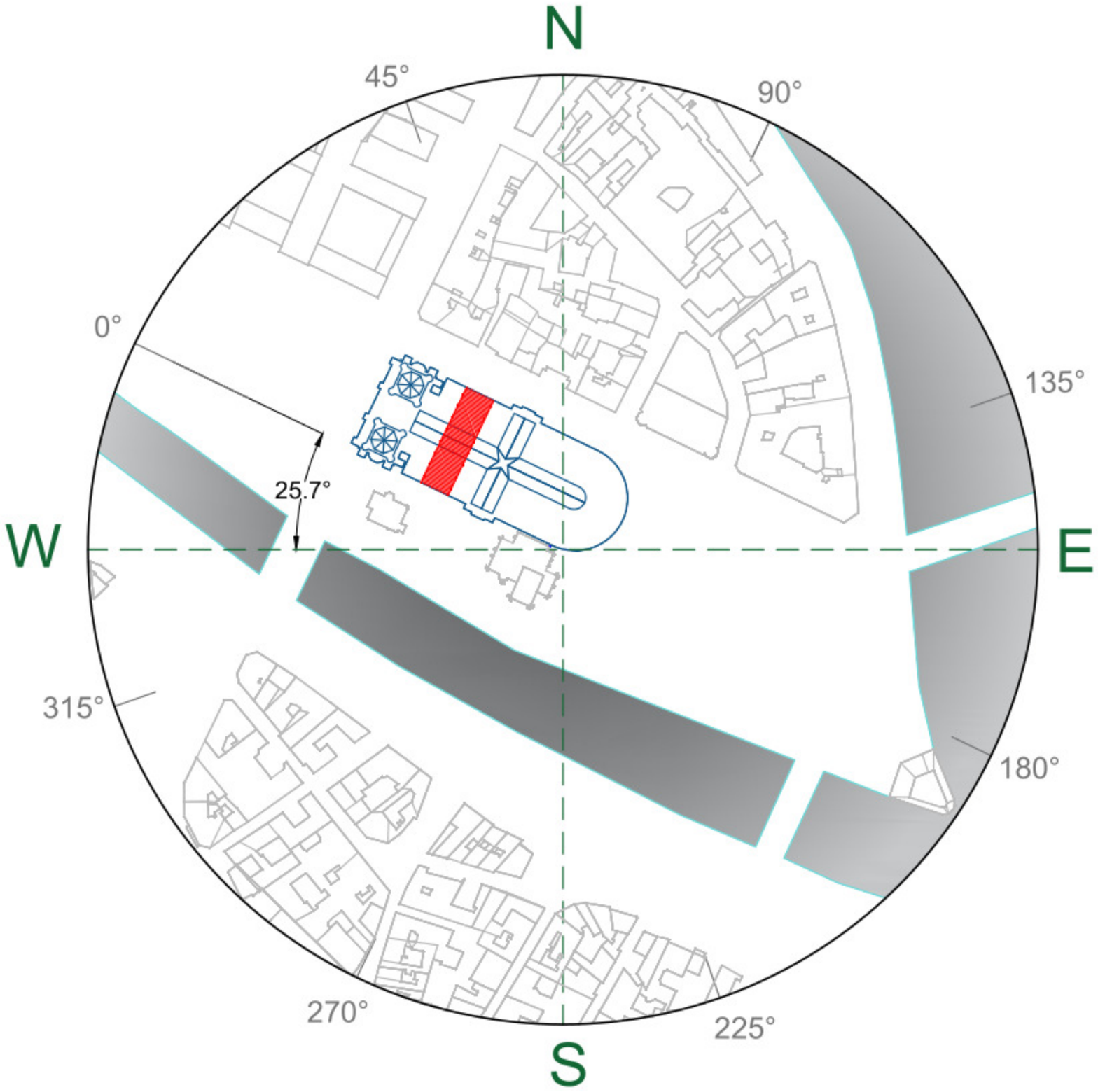}
	\caption{Scheme indicating the wind velocity directions; the portion of the Cathedral that has been studied in  detail is  highlighted in red.}
	\label{fig:5}
	\end{center}
\end{figure}

To determine the wind pressure field  over the whole external surface of the Cathedral, the physical model of Notre Dame has been equipped with  1200 pressure gauges, whose distribution has been studied in order to obtain, by interpolation methods, detailed charts of the pressure coefficients, cf. Sect. \ref{sec:results}. As previously said, the study has been conducted on the Cathedral model with and without the surrounding parts of Paris, see Fig. \ref{fig:4}.
This, for two reasons: on the one hand, to evaluate the influence of the surrounding buildings and, on the other hand, for obtaining pressure coefficient distributions that can represent the wind loading on similar buildings immersed in a generic urban wind profile (considering Notre Dame as a sort of Gothic Cathedral archetype).
In both cases, the wind profile has been generated through artificial roughness elements, cf. Sect.~\ref{sec:flow}. The test campaign has been carried out recording the pressure on the surface of the Cathedral for all wind directions, according to the scheme presented in Fig.~\ref{fig:5}. Therein, the orientation of the Cathedral with respect to the canonical geographical directions, the wind tunnel azimuths and the surrounding buildings can be seen. The direction denoted as 0$^\circ$ (West-Northwest) is perpendicular to the main façade, while 90$^\circ$ means that the wind blows perpendicularly to the naves from the side where the neighboring buildings are closer to the Cathedral (North-Northeast). A wind direction of 270$^\circ$ indicates a wind perpendicular to the naves but coming from the South-Southwest side.
%\begin{comment}

%\end{comment}
%\caption{Scheme indicating the wind velocity directions, left (the  portion of the Cathedral that has been studied in great details is also highlighted in red), and its positioning on the rotating board in the wind tunnel, right.}

\subsection{Wind tunnel facility}
\label{sec:wind_tunnel}

The tests were carried out in the open-circuit boundary layer wind tunnel of CRIACIV\footnote{Centro di Ricerca Interuniversitario di Aerodinamica delle Costruzioni e Ingegneria del Vento, Inter-University Research Centre on Building Aerodynamics and Wind Engineering} in Prato, Italy. The facility is 22~m long and presents at the inlet a nozzle with a contraction ratio of 3 to 1 after the honeycomb and a T-diffuser at the outlet. The test chamber is 1.6~m high, while the width varies from 2.2~m after the nozzle to 2.4~m at the position of the turning table. The latter has a diameter of 2.2~m. The overall length of the fetch to develop boundary layer flows is 11~m. Air is drawn by a motor with a nominal power of 156~kW, and the flow speed can be varied continuously up to about 30~m/s by adjusting through an inverter the rotation speed of the fan or the pitch of its ten blades. In the absence of turbulence generating devices, the free-stream turbulence intensity is less than 1\%.

\subsection{Model equipment}
\label{sec:model}
\begin{figure}[h!]
	\begin{center}
	\includegraphics[width=0.5\textwidth, angle=270]{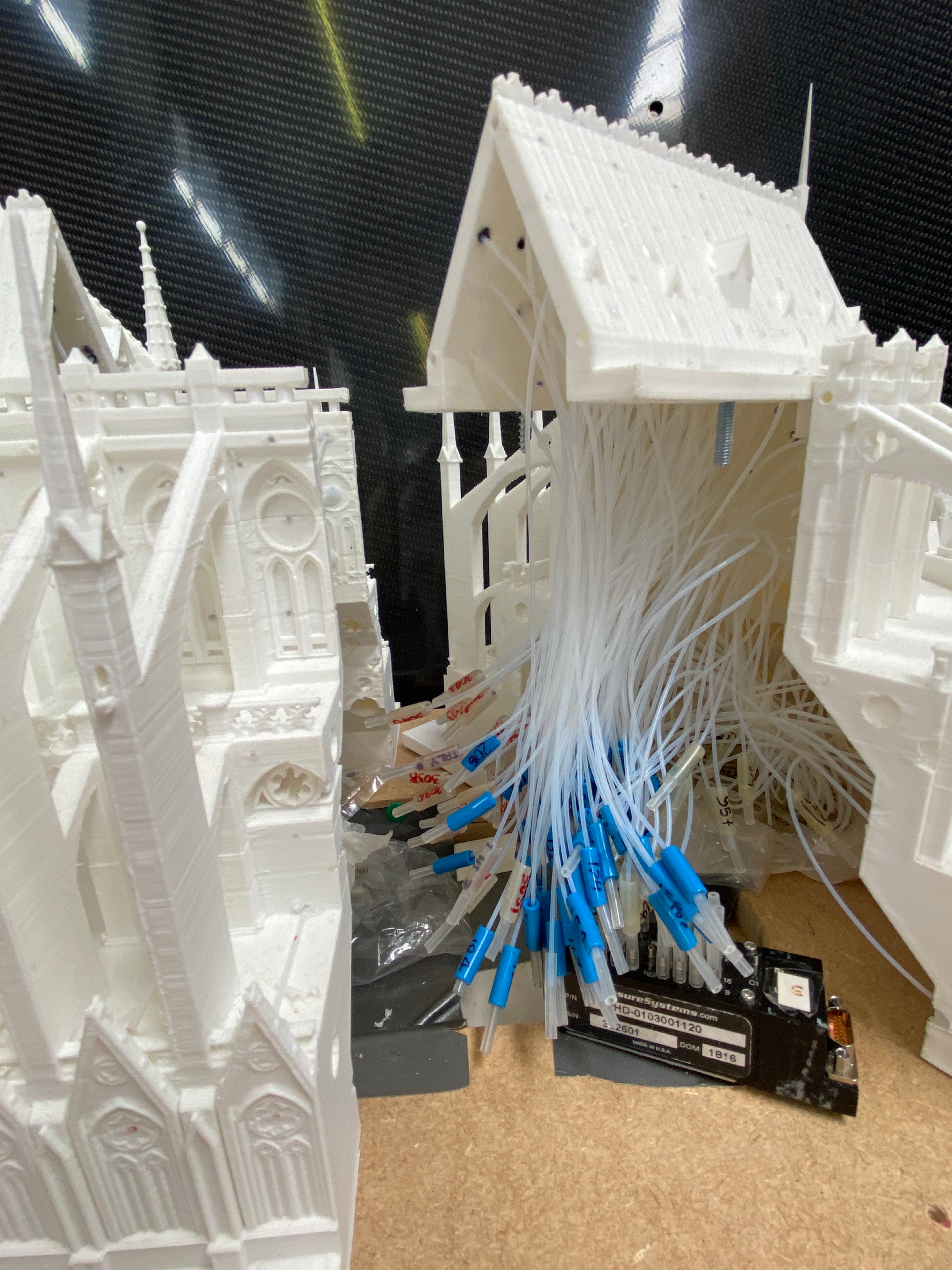}
	\caption{Parts of the model equipped with pressure taps; the Teflon tubes are connected to an ESP pressure scanner (the black box visible at the bottom).}
	\label{fig:6}
	\end{center}
\end{figure}

The model of the Cathedral has been equipped with Teflon tubes with an internal diameter of 1~mm, used to connect the taps on the model surface with the pressure sensors, see Fig.~\ref{fig:6}. The tubes were 30~cm long, obtained by linking two pieces having the same internal diameter with a short restrictor tube of about 1~cm working as a damper. Such pneumatic connections have been calibrated beforehand to guarantee a properly flat transfer function in the frequency range of interest (up to about 200~Hz in the present case).

%\begin{comment}
%\end{comment}

222 pressure taps were simultaneously recorded at a sampling rate of 500~Hz with the system PSI DTC Initium. The accuracy of the piezoelectric sensors of the 32-port miniaturized ESP pressure scanners is $\pm$2.45~Pa. Each signal was recorded for about 120~s at a reference mean wind speed of about 19.5~m/s at the top of the Cathedral roof.

At full scale, following the Eurocode~1, \citet{Eurocode1}, a mean wind speed of 21.3~m/s is expected at a height of 44~m (top of the roof) for a terrain category IV (urban profile) and a return period of 50 years. Consequently, the velocity scale of the tests is about 1:1.1, while the time scale results to be about 1:183.
%\begin{comment}
\begin{figure}[h!]
	\begin{center}
	\subfigure[\label{fig:U_profile}]
  {\includegraphics[angle=0, width=0.485\textwidth]{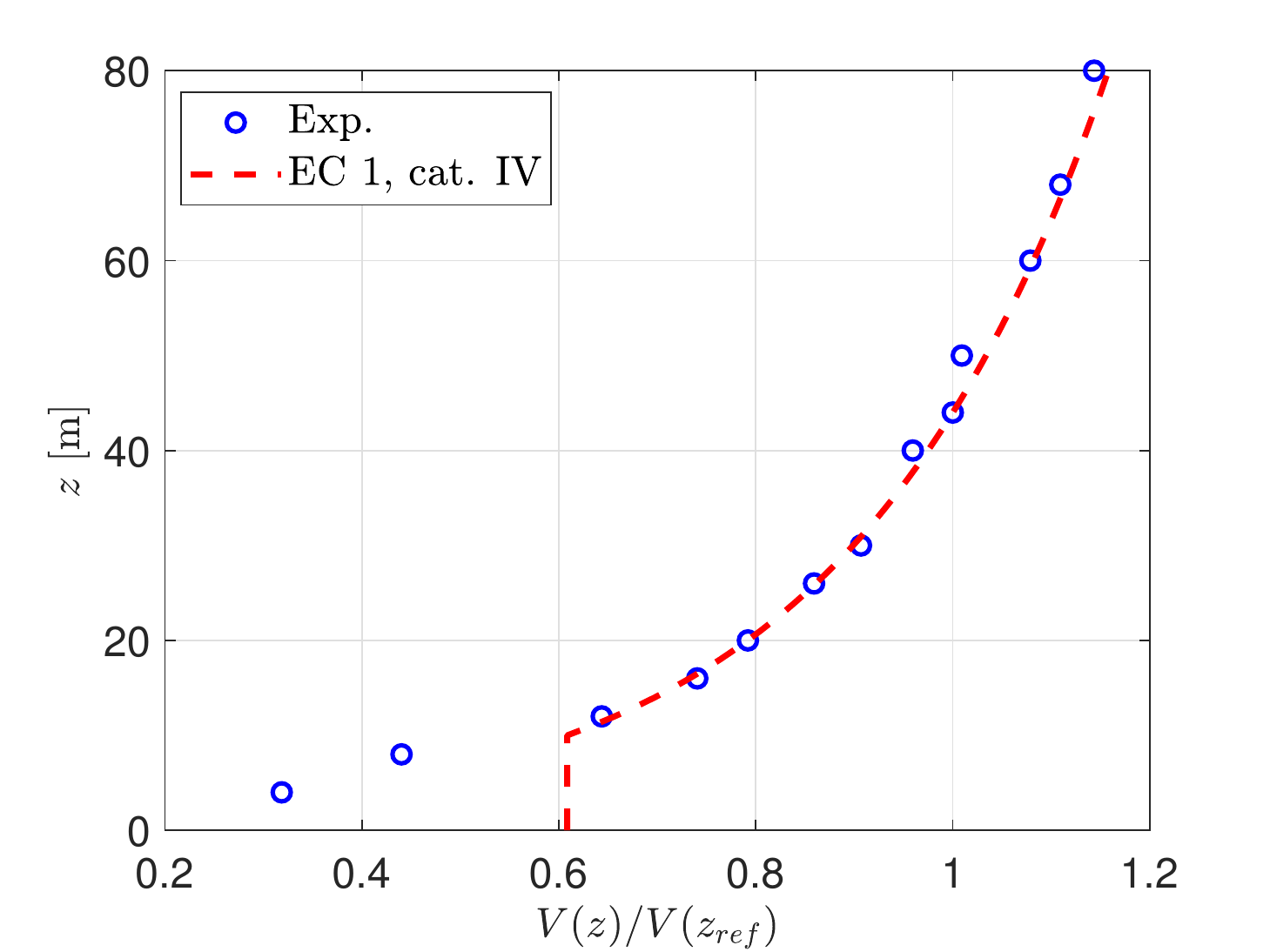}}
	\subfigure[\label{fig:Iu_profile}]
  {\includegraphics[angle=0, width=0.485\textwidth]{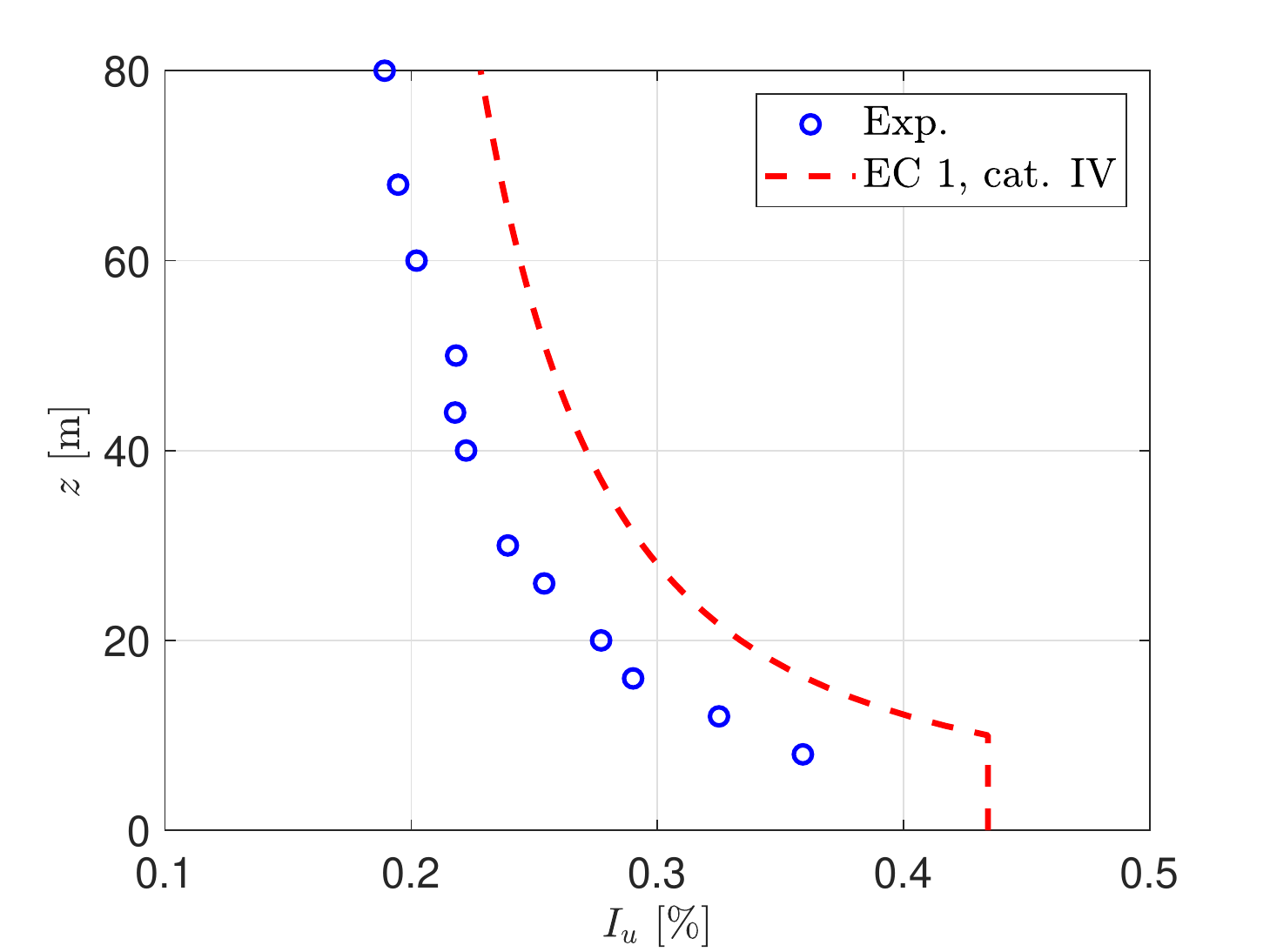}}\\
	\subfigure[\label{fig:Lux_profile}]
  {\includegraphics[angle=0, width=0.485\textwidth]{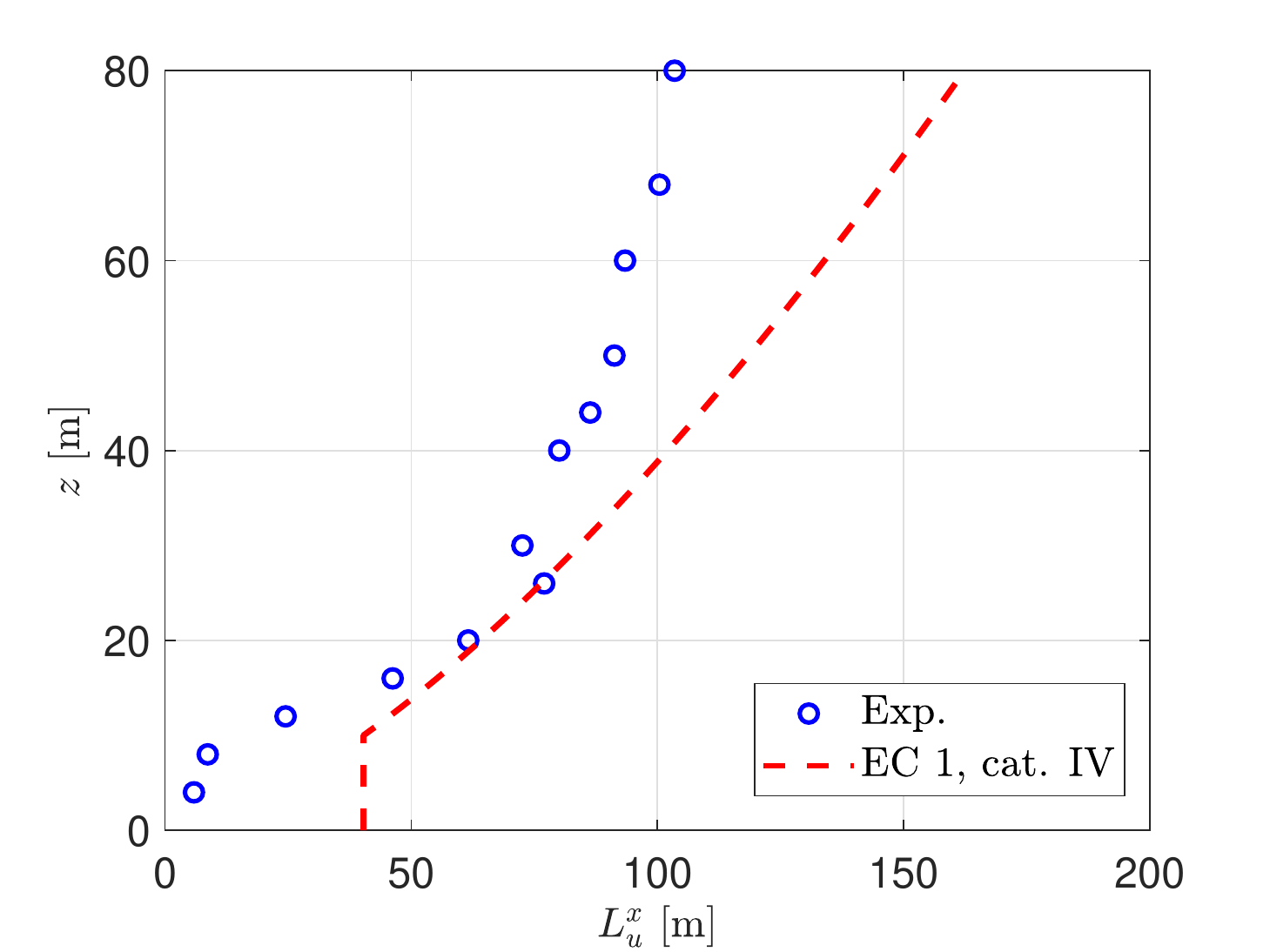}}
	\subfigure[\label{fig:wind_PSD}]
  {\includegraphics[angle=0, width=0.485\textwidth]{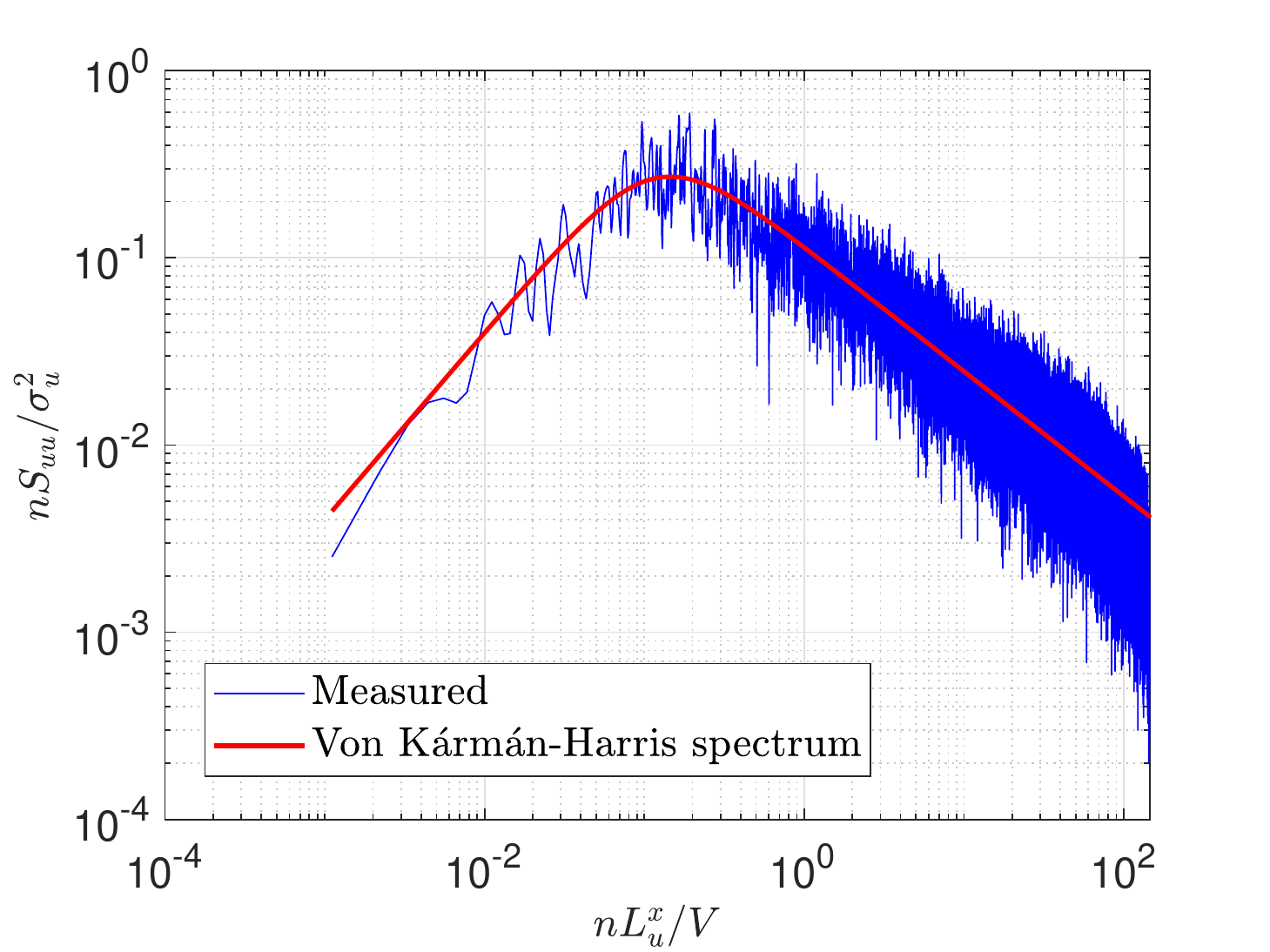}}
	\caption{(a) Normalized mean velocity ($z_{ref} = H = 44$~m denotes the reference height, i.e., the top of the roof), (b) turbulence intensity, and (c) longitudinal integral length scale of turbulence profiles in the present tests. The power spectral density of longitudinal velocity fluctuations at the reference height is shown too (d).}
	\label{fig:7}
	\end{center}
\end{figure}
%\end{comment}

\subsection{Oncoming flow}
\label{sec:flow}

For the city centre of Paris, the previously mentioned turbulent wind profile associated with the terrain category IV of Eurocode~1 has been assumed for all azimuthal directions. The atmospheric boundary layer flow at the same scale of the model (1:200) has been reproduced through artificial roughness elements of variable size diffused all over the floor of the test chamber. A castellated barrier, placed at the inlet section of the wind tunnel has also been employed to increase the turbulence intensity. 

Fig.~\ref{fig:7} shows the modeled wind characteristics in the wind tunnel at the beginning of the turning table, slightly upstream of the model of the Cathedral. Flow velocity measurements were carried out with a single-component hot-wire anemometer recorded at a sampling rate of 10~kHz. The mean wind velocity pattern is in very good agreement with the assumed target urban profile (Fig.~\ref{fig:U_profile}). The longitudinal turbulence intensity, which is a normalized integral measure of the wind velocity fluctuations, is slightly lower than the Eurocode~1 target (Fig.~\ref{fig:Iu_profile}). The longitudinal integral length scale of turbulence, which represents the correlation length of the wind velocity fluctuations along the mean velocity direction and rules the frequency distribution of the turbulent kinetic energy in the stream, closely follows the target pattern up to about 30~m above the ground; afterwards, the increase with the height becomes slower than in the Eurocode~1 profile (Fig.~\ref{fig:Iu_profile}). However, the discrepancy is moderate at least up to the top of the Cathedral roof. Finally, Fig.~\ref{fig:wind_PSD} shows that the spectral characteristics of the generated boundary layer comply very well with a von K\'arm\'an-Harris spectrum, which is very often assumed to describe the energy cascade of turbulence, \citet{Simiu_Yeo}. Measurements have also been carried out to assess the homogeneity of the flow in the transversal and longitudinal directions.

In general, despite the large scale of the model with respect to the present wind tunnel facility, we can conclude that the modeled turbulent wind is well representative of an urban environment such as the one that can be expected in the centre of Paris.

As said above, two different configurations of the model have been tested. In the first one, the turning table was covered with roughness elements progressively degrading close to the Cathedral model, cf. Fig.~\ref{fig:4} a), in order to transfer the same generic urban wind profile from the beginning of the turning table to the model position and beyond. In contrast, the second configuration includes the model of the surrounding of the Cathedral, as shown in Fig.~\ref{fig:4} b).
%\begin{comment}
\begin{figure}[h!]
	\begin{center}
	\subfigure[\label{fig:Fiancata_windward_medie_surr}]
  {\includegraphics[width=\textwidth]{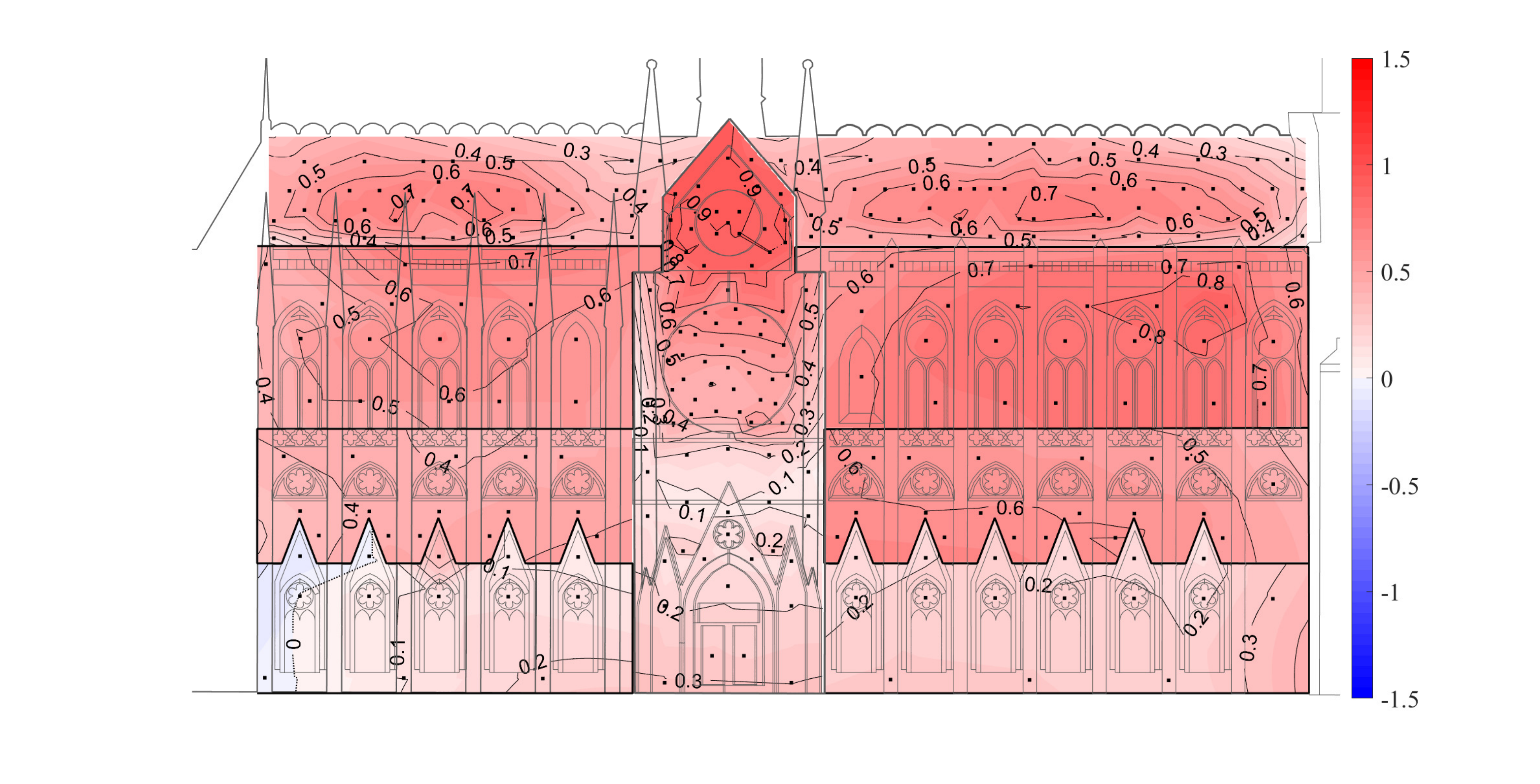}}\\
	\subfigure[\label{fig:Fiancata_leeward_medie_surr}]
  {\includegraphics[width=\textwidth]{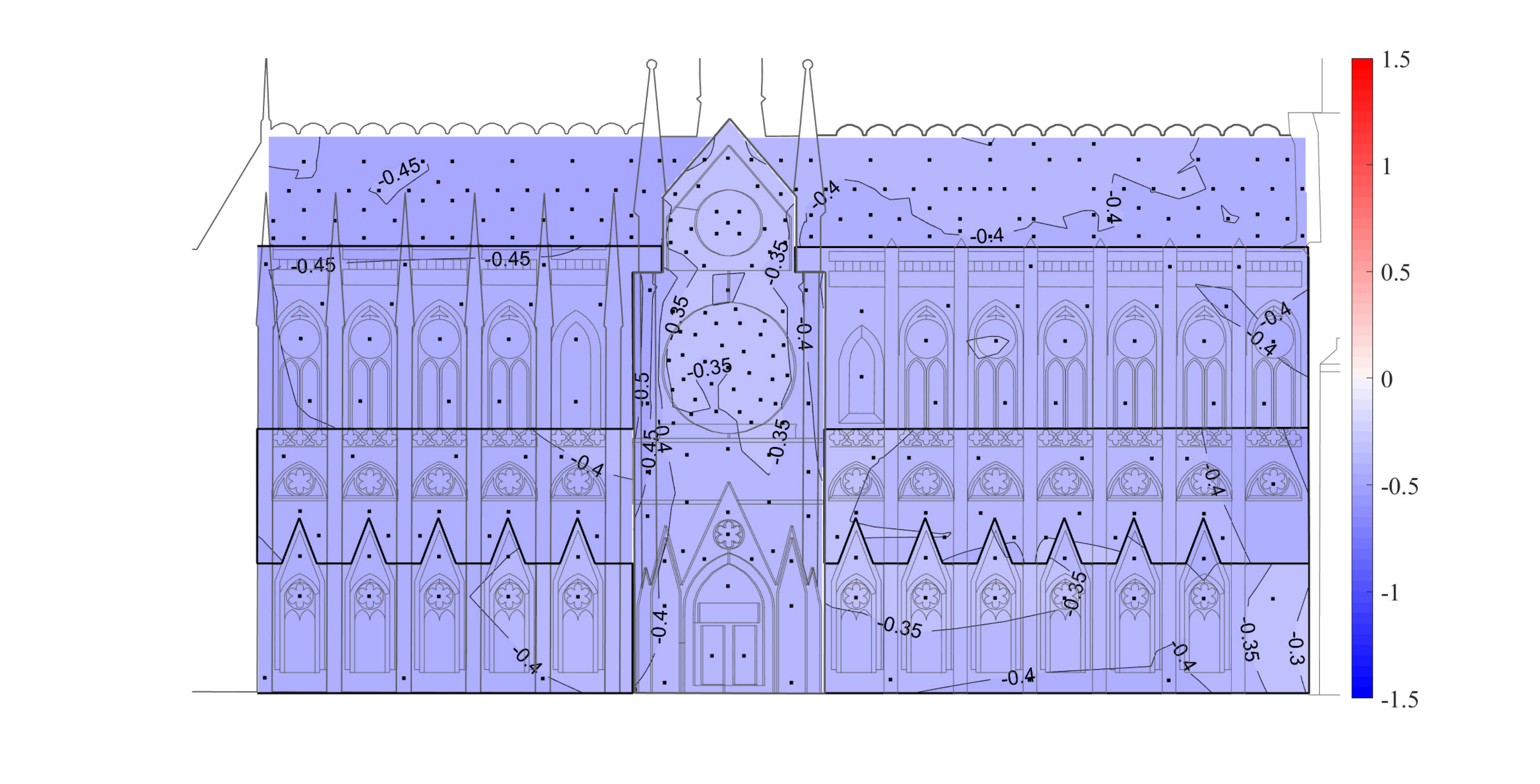}}
	\caption{Mean pressure coefficient distribution on the flank of the Cathedral for the configuration with surrounding: windward side (a) and leeward side (b). The wind velocity direction is North-Northeast, perpendicular to the side walls and denoted as 90$^\circ$ in Fig.~\ref{fig:5}.}
		\label{fig:8}
	\end{center}
\end{figure}
%\end{comment}

\section{Results of the experimental tests}
\label{sec:results}

\subsection{Recorded measurements}

The results of pressure measurements on the external surface of the Cathedral are reported in this section in the form of pressure coefficients, defined as follows:
\begin{equation}
 C_p = \frac{p - p_0}{\frac{1}{2}\rho V^2(z_{ref})},
\label{eq:Cp}
\end{equation}
where $p$ is the static pressure on the considered point of the structure, $p_0$ is the static pressure of the undisturbed flow in the wind tunnel at the position of the model, $\rho$ is the air density and $V(z_{ref})$ is the reference wind speed, that is the mean flow velocity at the top of the roof ($z_{ref} = H = 44$~m at full scale) in the absence of the model (see the mean wind velocity profile in Fig.~\ref{fig:U_profile}).
It is noteworthy that the accuracy of the pressure transducers, given the reference velocity pressure in the tests, corresponds to $\pm 0.01$ in terms of $C_p$. The recorded data can be divided in mean pressure and gust pressure data. The former are simply calculated as the time averages of the recorded pressure coefficients. The latter are defined as the average of either the maximum or the minimum values of $C_p$ registered over full-scale time windows of 10 minutes (see Section~\ref{sec:peak_loads}).
The results are reported hereafter in terms of pressure coefficient charts, obtained by linear interpolation and heuristically controlled extrapolation of the measured values. The small black spots indicate the taps where pressure was actually measured.

\subsection{Mean pressure charts}
\label{mean_charts}

Fig.~\ref{fig:8} gives the global distribution of the pressure coefficient on the lateral side of the Cathedral in the presence of the surrounding model and wind coming from the 90$^\circ$-direction. It is worth noting the strong pressure gradient in correspondence of the windward rose of the transept and the large pressure coefficients in the higher part of the transept ($C_p > 0.9$).

\begin{figure}[h!]
	\begin{center}
  \includegraphics[width=1.0\textwidth]{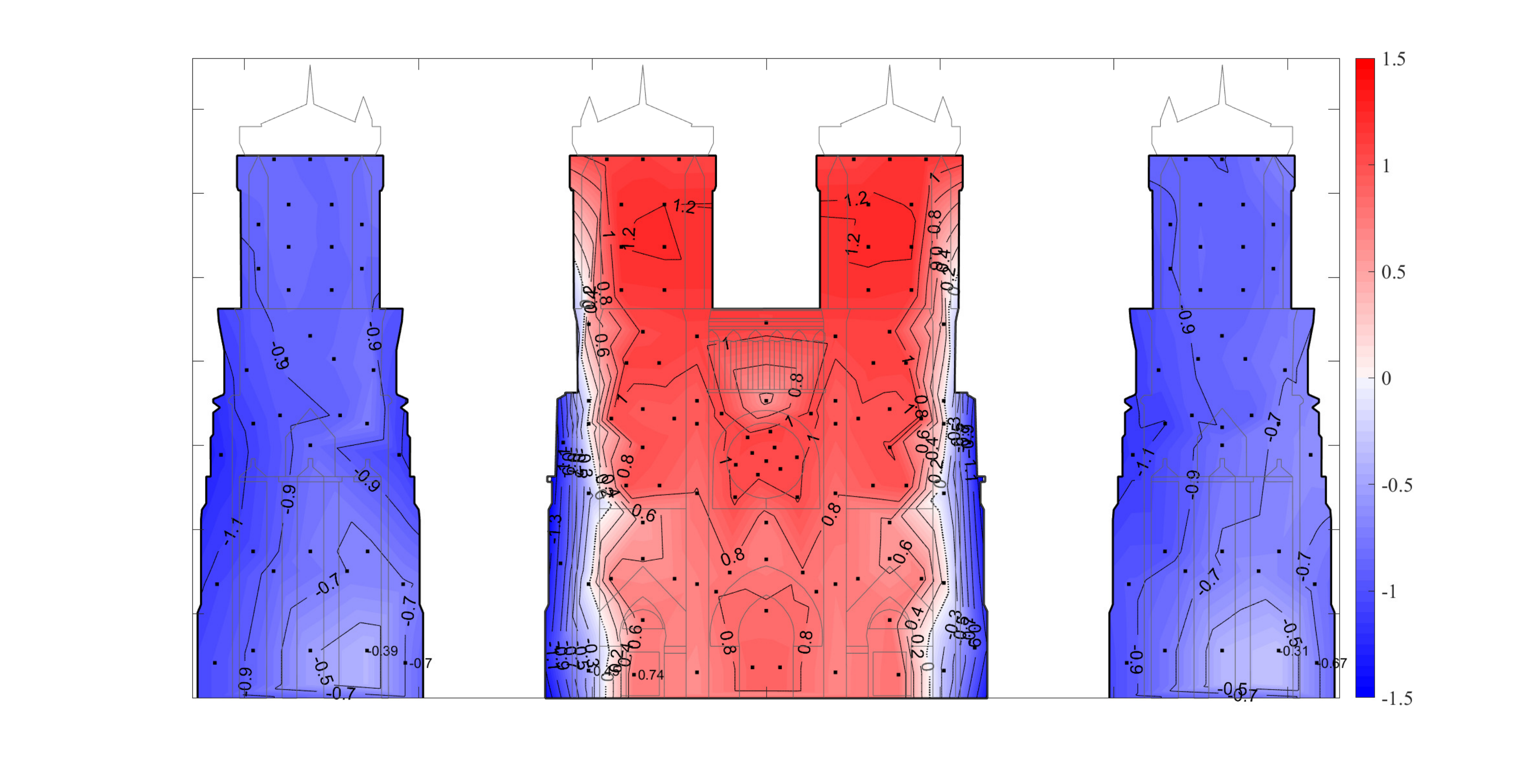}
	\caption{Mean pressure coefficient distribution on the front of the Cathedral and on the external sides of the towers for the configuration with surrounding. The wind velocity direction is West-Northwest, perpendicular to the main façade and denoted as 0$^\circ$ in Fig.~\ref{fig:5}.}
		\label{fig:façade}
	\end{center}
\end{figure}

The pressure coefficient distribution over the front of the Cathedral is reported in Fig.~\ref{fig:façade} for a wind perpendicular to it (0$^\circ$-direction). It is apparent that the particular geometry of the towers produces large pressure gradients close to the edges of the façade. It is also noteworthy that values of $C_p$ larger than unity can be found in the upper part of the towers; this is because the reference velocity pressure is taken at a lower height, that is that of the top of the roof (see Eq.~\ref{eq:Cp}).

\begin{figure}[h!]
	\begin{center}
	\subfigure[\label{fig:Apse_90deg}]
  {\includegraphics[width=\textwidth]{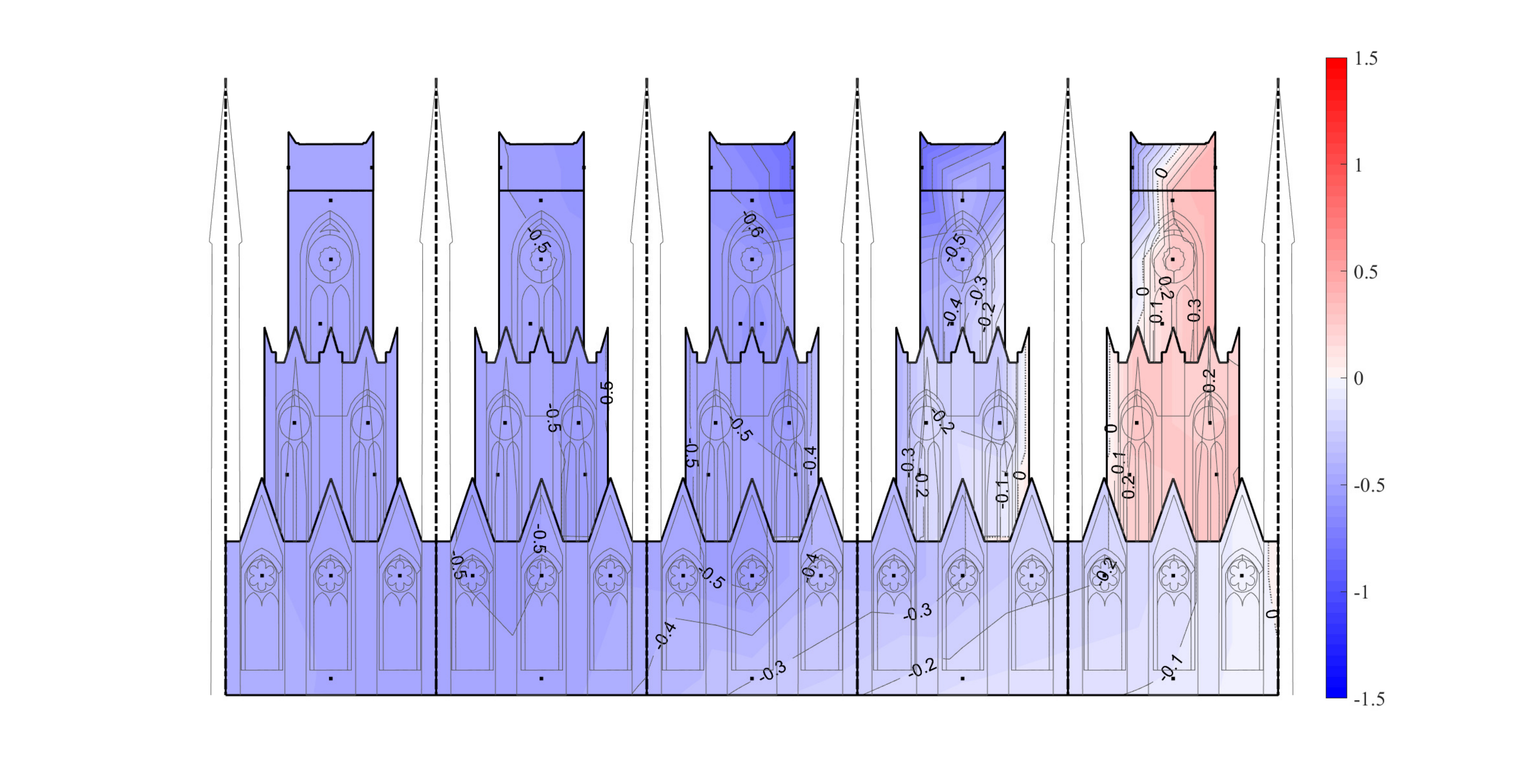}}\\
	\subfigure[\label{fig:Apse_180deg}]
  {\includegraphics[width=\textwidth]{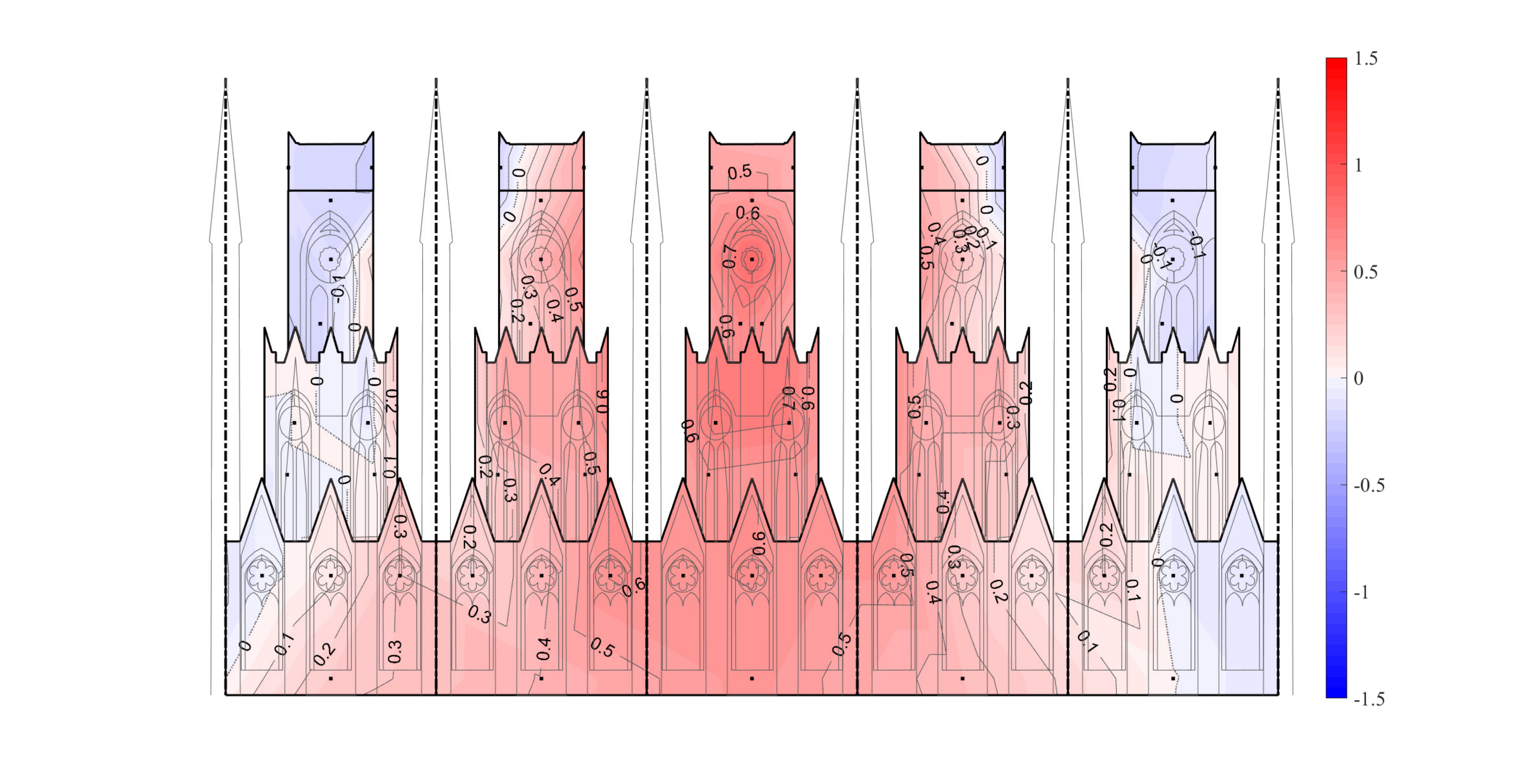}}
	\caption{Mean pressure coefficient distribution on the development of the apse of the Cathedral for the configuration with surrounding: wind velocity directions of 90$^\circ$ (a) and 180$^\circ$ (b).}
		\label{fig:Apse}
	\end{center}
\end{figure}

Mean pressure coefficient charts are presented in Fig.~\ref{fig:Apse} also for the apse, for a wind blowing either from the direction perpendicular to the left flank (90$^\circ$) or parallel to the longitudinal axis of the Cathedral (180$^\circ$). In the former case, one can notice a strong pressure gradient due to the curvature of the walls, which from positive pressures on the windward side quickly leads to strong suctions in the middle upper part. In contrast, for a wind direction of 180$^\circ$ positive mean pressure coefficients up to about 0.8 are attained in the central upper portion of the apse.
In the lower part of the left side of the apse (bottom right corner in Fig.~\ref{fig:Apse_90deg}-\ref{fig:Apse_180deg}), one can also remark the effect of the surrounding buildings, which are very close to the apse. They shelter the flank of the Cathedral from direct wind for an azimuth angle of 90$^\circ$, while they promote a flow acceleration, and then a pressure decrease, for a wind direction of 180$^\circ$.
%\begin{comment}

%\end{comment}

\subsubsection{Detail of the flank}
\label{fascia_2D}

\begin{figure}[t]
	\begin{center}
	\subfigure[\label{fig:Fascia_2D_no_surr}]
  {\includegraphics[width=0.5\textwidth]{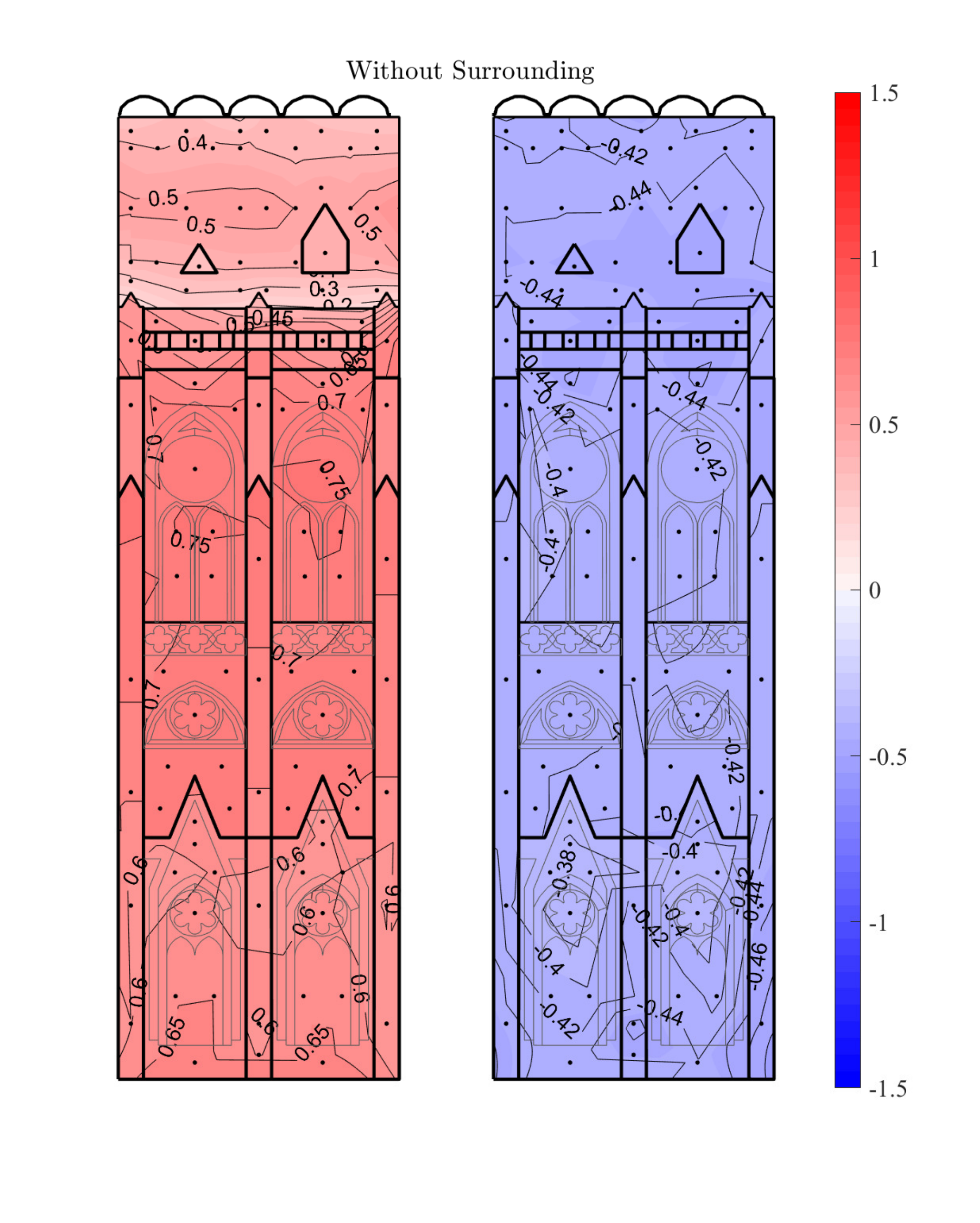}}%\hspace{-1.5cm}
	\subfigure[\label{fig:Fascia_2D_surr}]
  {\includegraphics[width=0.5\textwidth]{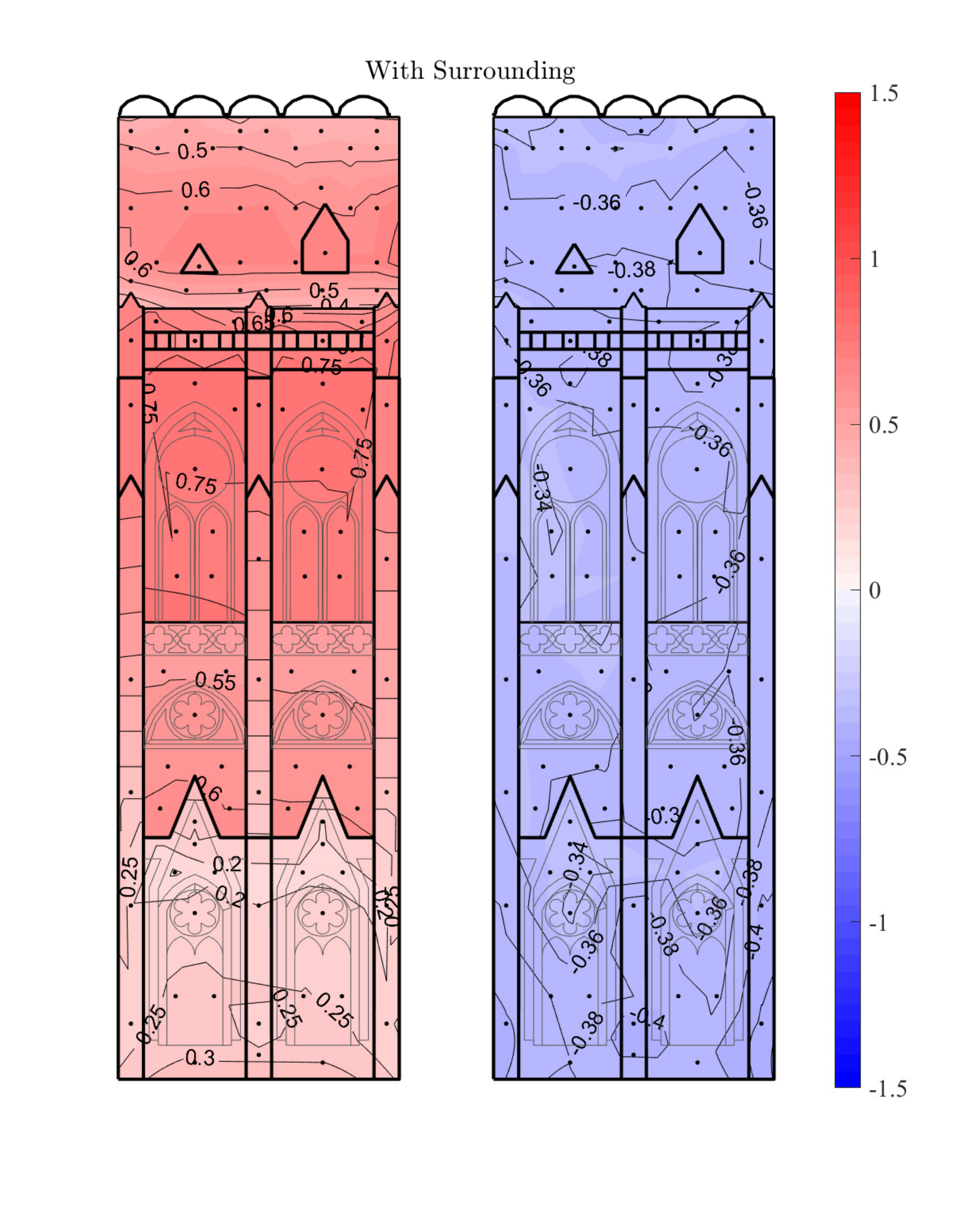}}
	\caption{Mean pressure coefficient distribution on a lateral portion of the Cathedral (see Fig.~\ref{fig:5}) for the configuration without the model of the surrounding (a) and including the buildings around the Cathedral (b). Each frame reports on the left the windward side and on the right the leeward side. The wind velocity direction is that perpendicular to the North-Northeast flank, denoted as 90$^\circ$ in Fig.~\ref{fig:5}.}
	\label{fig:10}
	\end{center}
\end{figure}

\begin{figure}[h!]
	\begin{center}
  \includegraphics[width=0.7\textwidth]{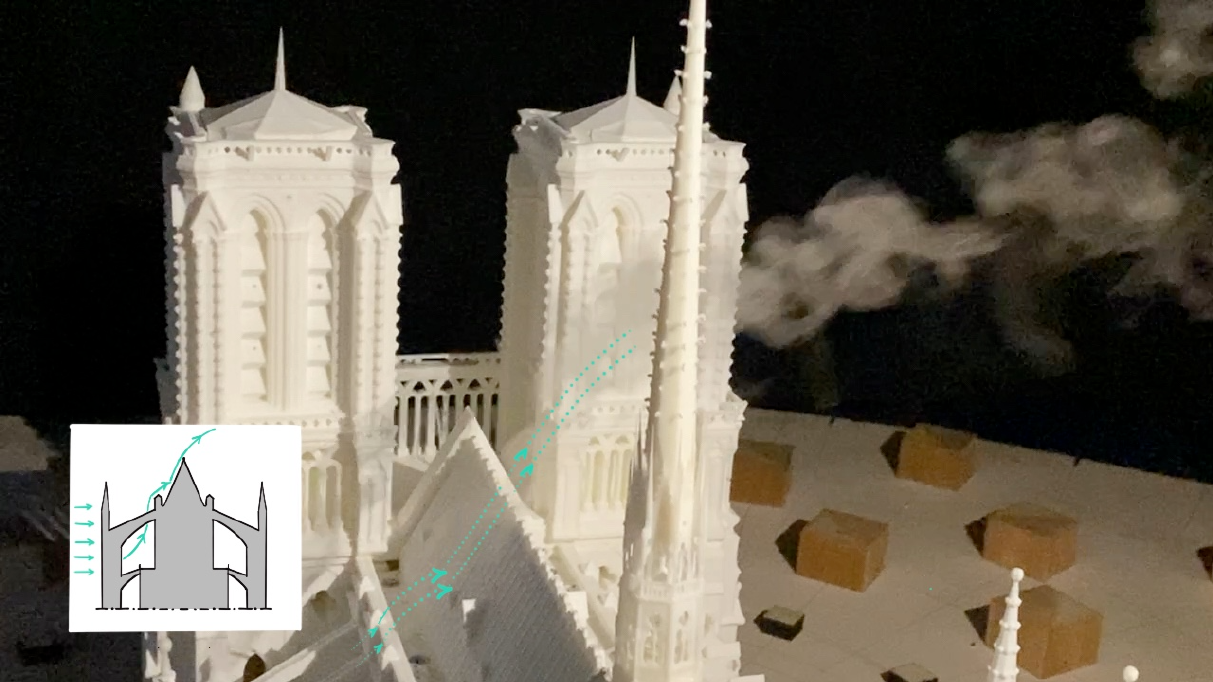}
	\caption{Flow visualization with a smoke generator in the roof region. The wind direction is perpendicular to the flank of the Cathedral (90$^\circ$).}
		\label{fig:fumo}
	\end{center}
\end{figure}

A lateral portion of the Cathedral between the front and the transept (indicated in red in Fig.~\ref{fig:5}) has been very densely instrumented, in order to have more details of the pressure coefficient distribution through the height of the Cathedral on a representative part of it. Results are shown in Fig.~\ref{fig:10}. In the case of a generic urban profile (Fig.~\ref{fig:Fascia_2D_no_surr}), pressure coefficients between about 0.6 and 0.75 are found on the windward walls. However, the most interesting features can be observed on the windward side of the roof, where a bubble of high pressure (with mean pressure coefficients up to about 0.55) is apparent in the central part. This is due to the flow acceleration on the upper side of the roof (with consequent decrease of pressure) and the strong inclination of the flow that locally separates at the balustrade at the base of the roof. Flow visualizations with a smoke generator (Fig.~\ref{fig:fumo}) revealed that this is also due to the significant vertical wind component induced by the three orders of walls on the flank of the Cathedral.
On the leeward side, the pressure is nearly uniform, assuming mean $C_p$ values between about $-0.45$ and $-0.4$.
%\begin{comment}

%\end{comment}
Fig.~\ref{fig:Fascia_2D_surr} clearly shows how the presence of the buildings upstream of the Cathedral and very close to it affects the load distribution. The pressure remarkably decreases on the lower part of the windward walls, but the vertical wind velocity component also reduces, producing a slight increase of the mean $C_p$ on the upper portion of the windward wall and a pronounced increment on the roof. A slight increase of pressure can also be observed on the leeward side of the Cathedral (lower suction).

\subsubsection{Roof}

\begin{figure}[h!]
	\begin{center}
	\subfigure[\label{fig:Tetto_medie_no_surr}]
  {\includegraphics[angle=0, width=\textwidth]{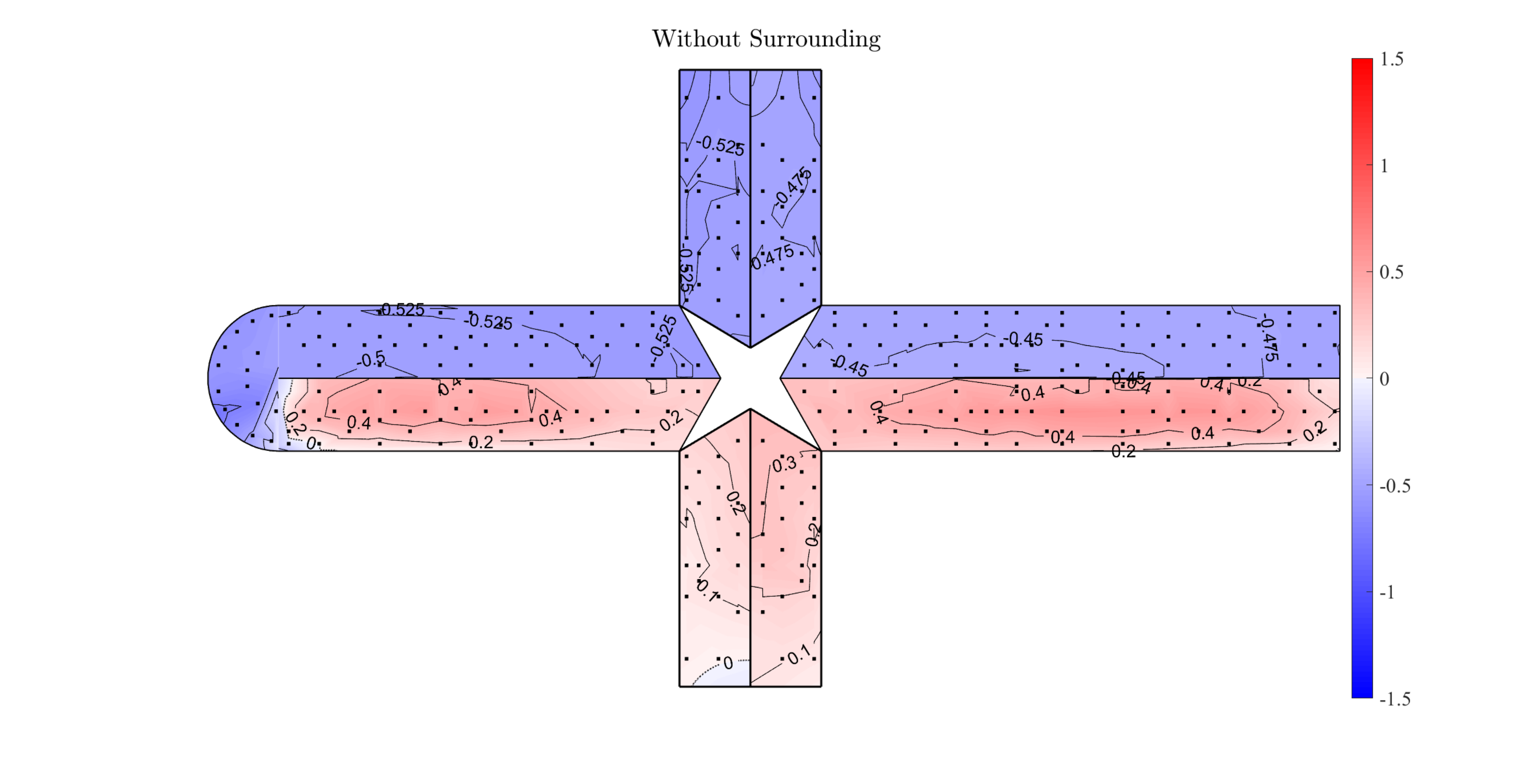}}\\
	\subfigure[\label{fig:Tetto_medie_surr}]
  {\includegraphics[angle=0, width=\textwidth]{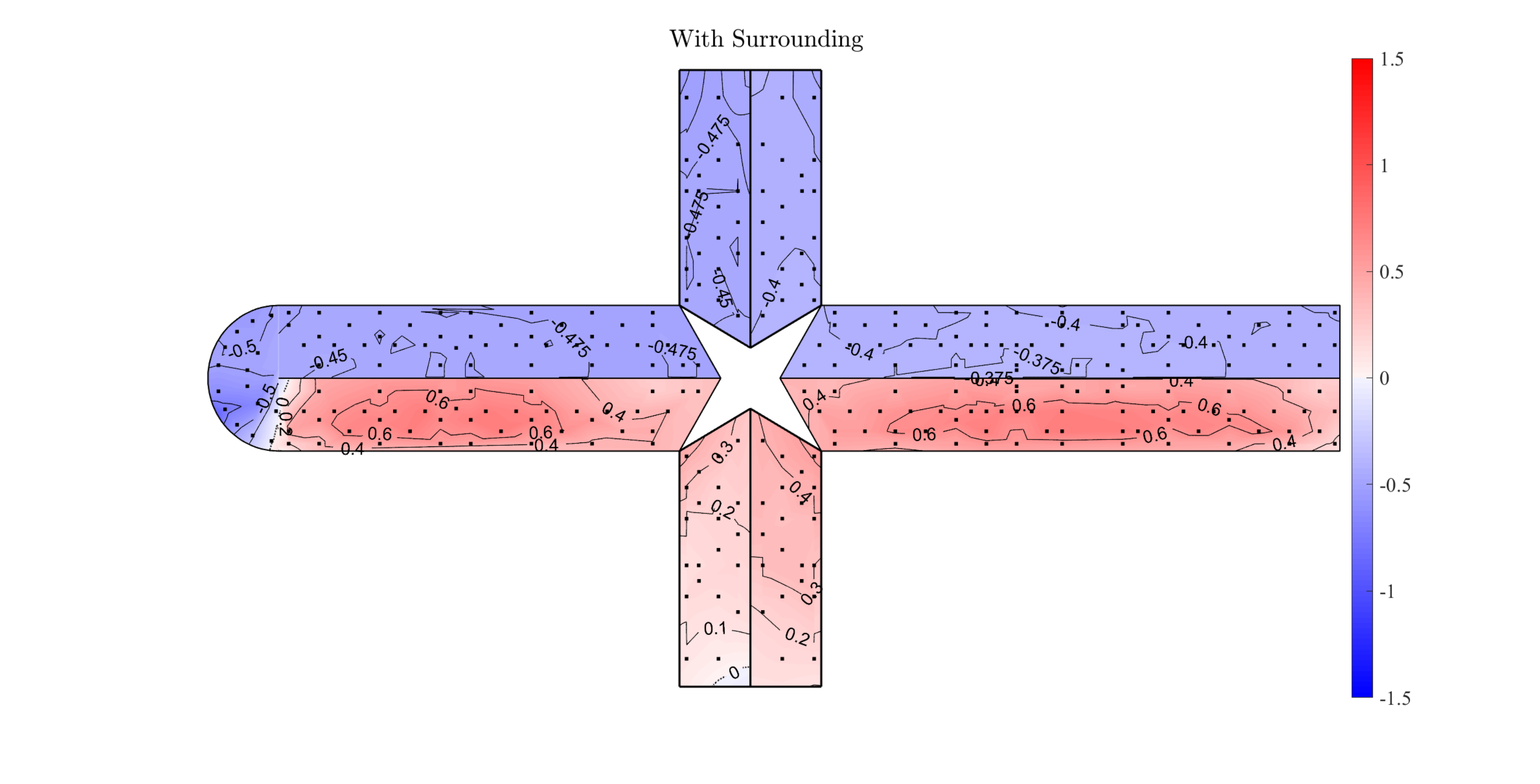}}
	\caption{Plan view of the mean pressure coefficients on the roof of the Cathedral: results without surrounding (a) and with surrounding (b). The wind velocity direction is that denoted as 90$^\circ$ in Fig.~\ref{fig:5}.}
	\label{fig:11}
	\end{center}
\end{figure}

The results discussed in the previous section are confirmed by Fig.~\ref{fig:11}, which reports the pressure distribution on the entire roof of the Cathedral. It can be remarked that, despite the roof is at a significantly greater height than the surrounding buildings, the presence of the latter has a remarkable impact on the pressure distribution over it, probably accelerating the upper flow and reducing the vertical component of the velocity with which the wind attacks the windward pitch of the roof. Indeed, while the pressure coefficient hardly attains 0.57 in a small windward region between the façade and the transept when the surrounding buildings are not reproduced, it overcomes 0.7 at several locations if the latter are in place.
%\begin{comment}

\begin{figure}[h!]
	\begin{center}
	\subfigure[\label{fig:Fiancata_windward_Max_No_surr}]
  {\includegraphics[width=\textwidth]{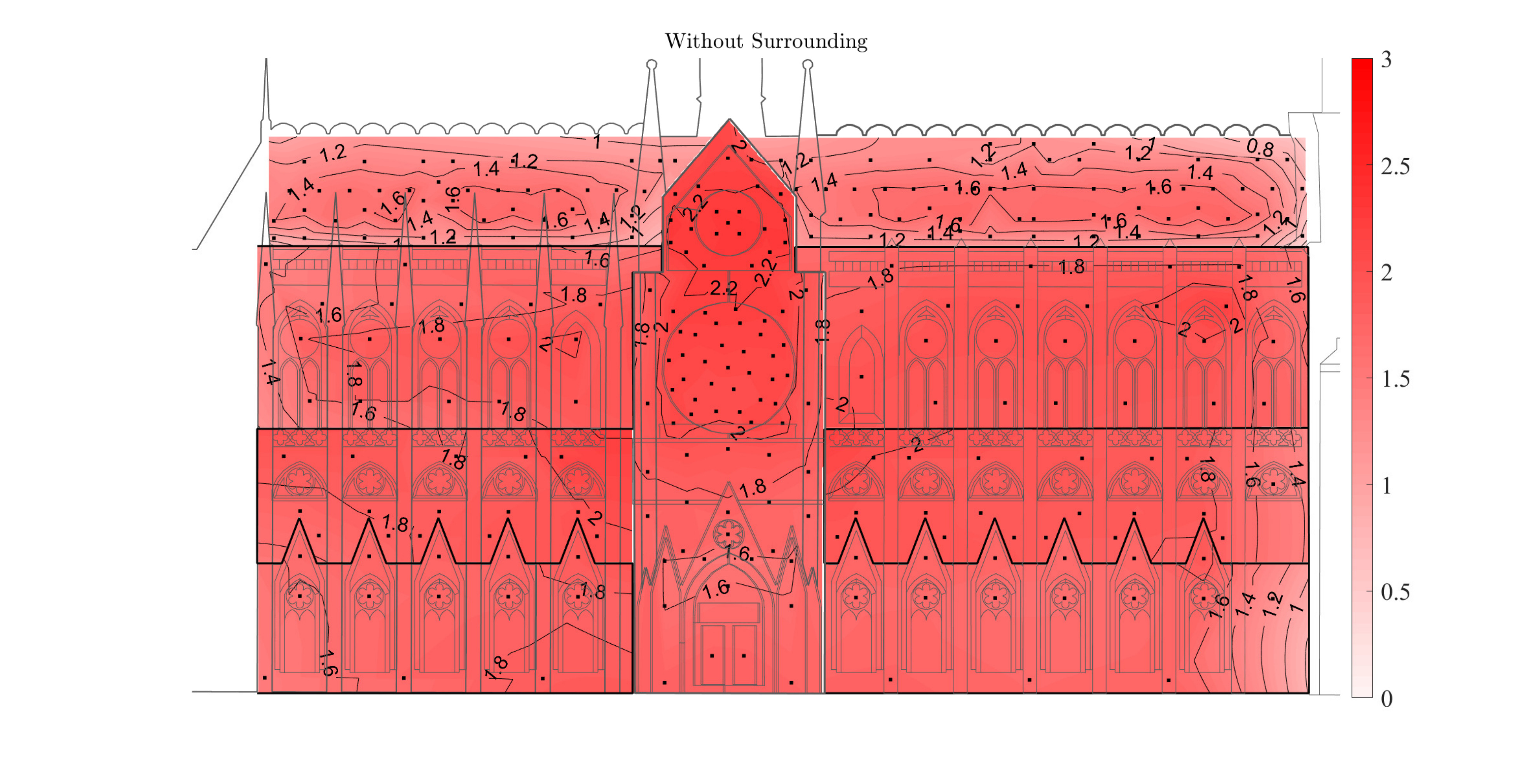}}\\
	\subfigure[\label{fig:Fiancata_windward_Max_Surr}]
  {\includegraphics[width=\textwidth]{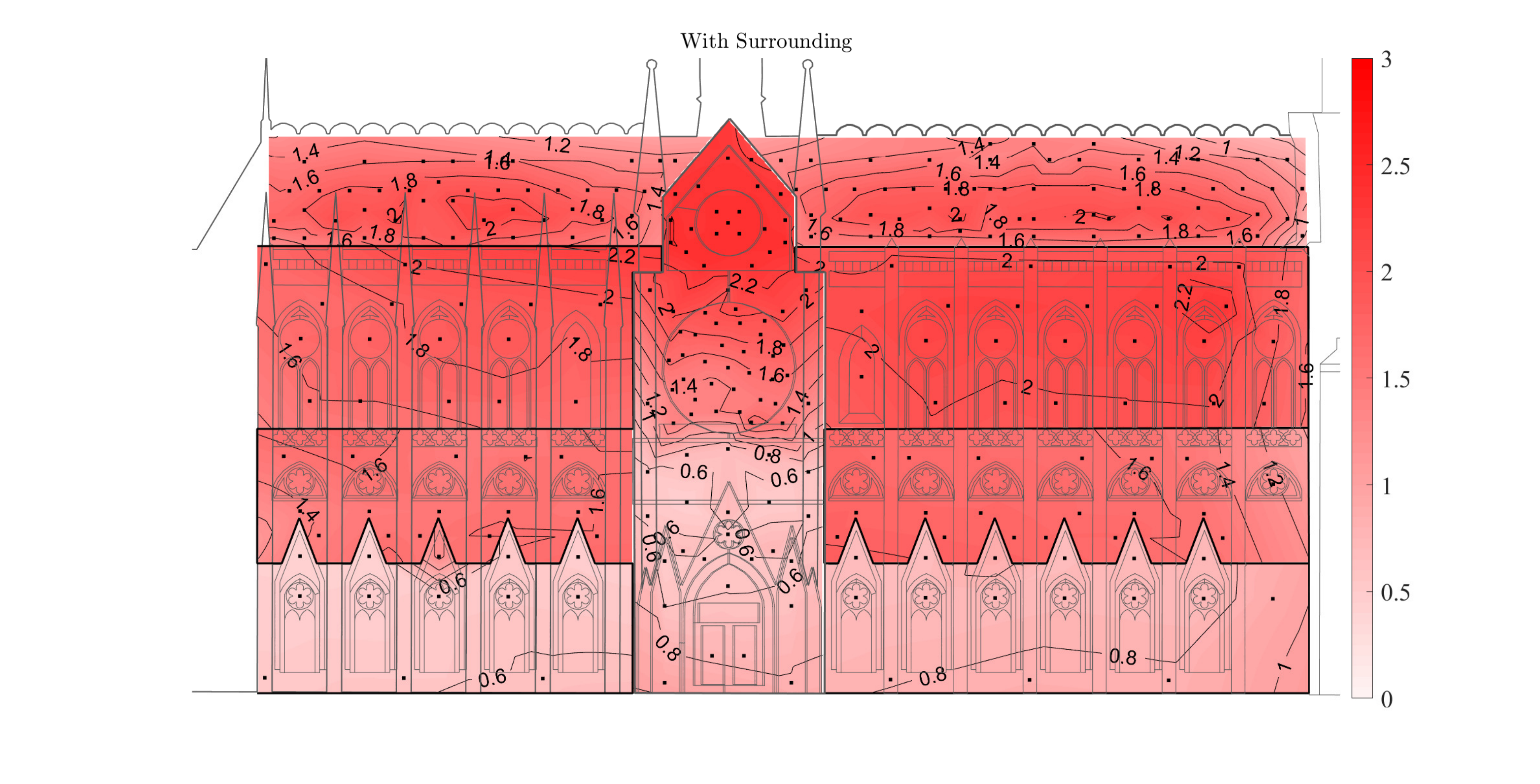}}
	\caption{Distribution of the maximum pressure coefficient on the windward flank of the Cathedral for a wind direction perpendicular to it (90$^\circ$): configuration without surrounding (a) and with surrounding (b).}
		\label{fig:Maxima_flank}
	\end{center}
\end{figure}

%\end{comment}

\subsection{Peak loads}
\label{sec:peak_loads}

For the local design or verification of cladding elements or secondary structures, and also to have a statistical measure of the fluctuation of pressures, peak values can be calculated according to Davenport approach, \citet{Davenport1961}, i.e. as the average of maxima (or minima) associated with a given observation time (600~s according to Eurocode~1's approach). Given the time scale of the current experiments, the length of the recorded signals corresponds to about 35 windows of 10 minutes at full scale and allows a statistically meaningful calculation of wind load peak values. Moreover, to be representative of area-averaged pressures, the maxima (or minima) are calculated on time records filtered via a moving average over a set time window, \citet{Lawson1976,Holmes1997}. In the present analysis, the classical full-scale window of 1~s has been chosen.

Fig.~\ref{fig:Maxima_flank} shows the distribution of the maximum pressure coefficients on the North-Northeast flank of the Cathedral for a wind perpendicular to the walls (90$^\circ$-direction). Large values of the load can be observed due to wind gusts, the $C_p$ chart showing values even beyond 2. The higher peak pressures are obtained just above the big rose window of the transept façade. The comparison with Fig.~\ref{fig:Fiancata_windward_medie_surr} shows that high gust factors (ratio of maximum to mean value of pressures) mostly between about 2.4 and 4 are obtained.

The comparison of Fig.~\ref{fig:Fiancata_windward_Max_No_surr} and Fig.~\ref{fig:Fiancata_windward_Max_Surr} emphasizes the effect of the surrounding buildings around the Cathedral on the peak pressure coefficients. Coherently with the observations already done for the mean pressures (Figs.~\ref{fig:10} and \ref{fig:11}), the presence of the surrounding strongly reduces the peak pressure coefficients on the lower part of the Cathedral flank, slightly increases them in the upper part of the lateral walls and on the transept façade, and remarkably intensifies the load on the roof.

Though not reported in Fig.~\ref{fig:Maxima_flank}, it is worth noting that the pressure fluctuations are such that the mimimum coefficients on the windward side of the Cathedral often reach negative values (nearly everywhere in the configuration with surrounding buildings).
%\begin{comment}
\begin{figure}[b!]
	\begin{center}
	\subfigure[\label{fig:Rosone_NO_Surr_windward}]
  {\includegraphics[width=0.485\textwidth]{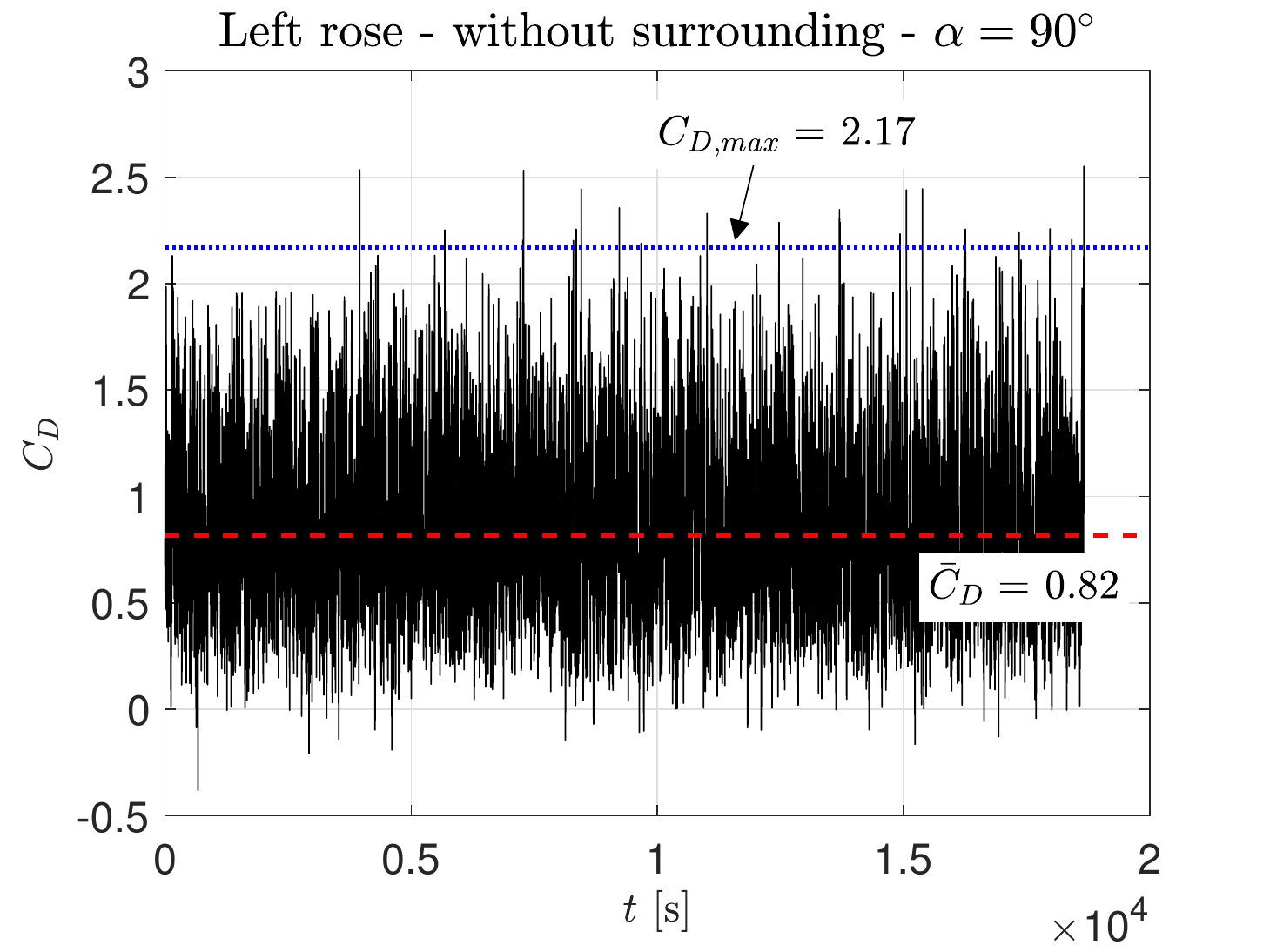}}
	\subfigure[\label{fig:Rosone_NO_Surr_leeward}]
  {\includegraphics[width=0.485\textwidth]{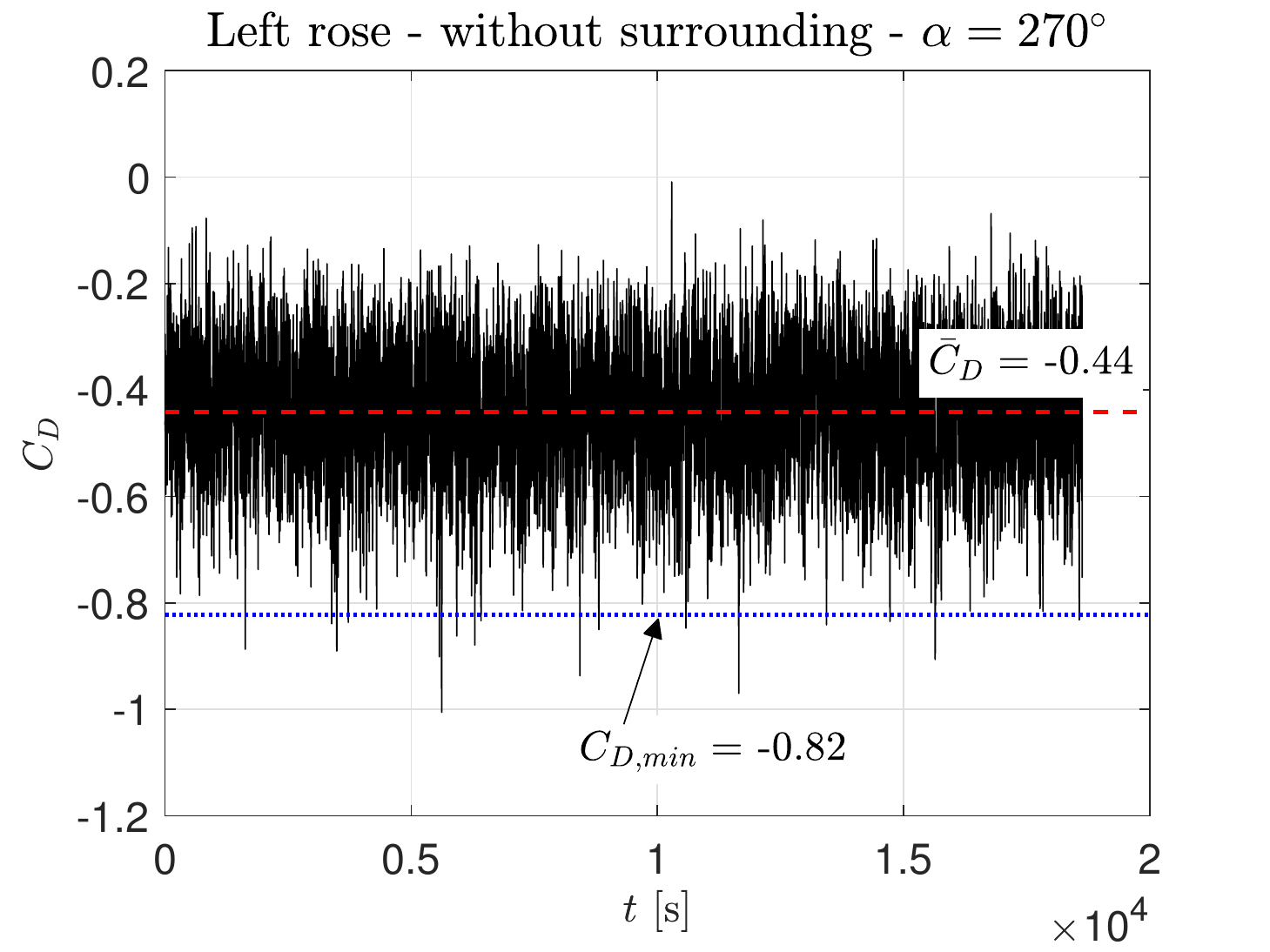}}\\
	\subfigure[\label{fig:Rosone_Surr_SX}]
  {\includegraphics[width=0.485\textwidth]{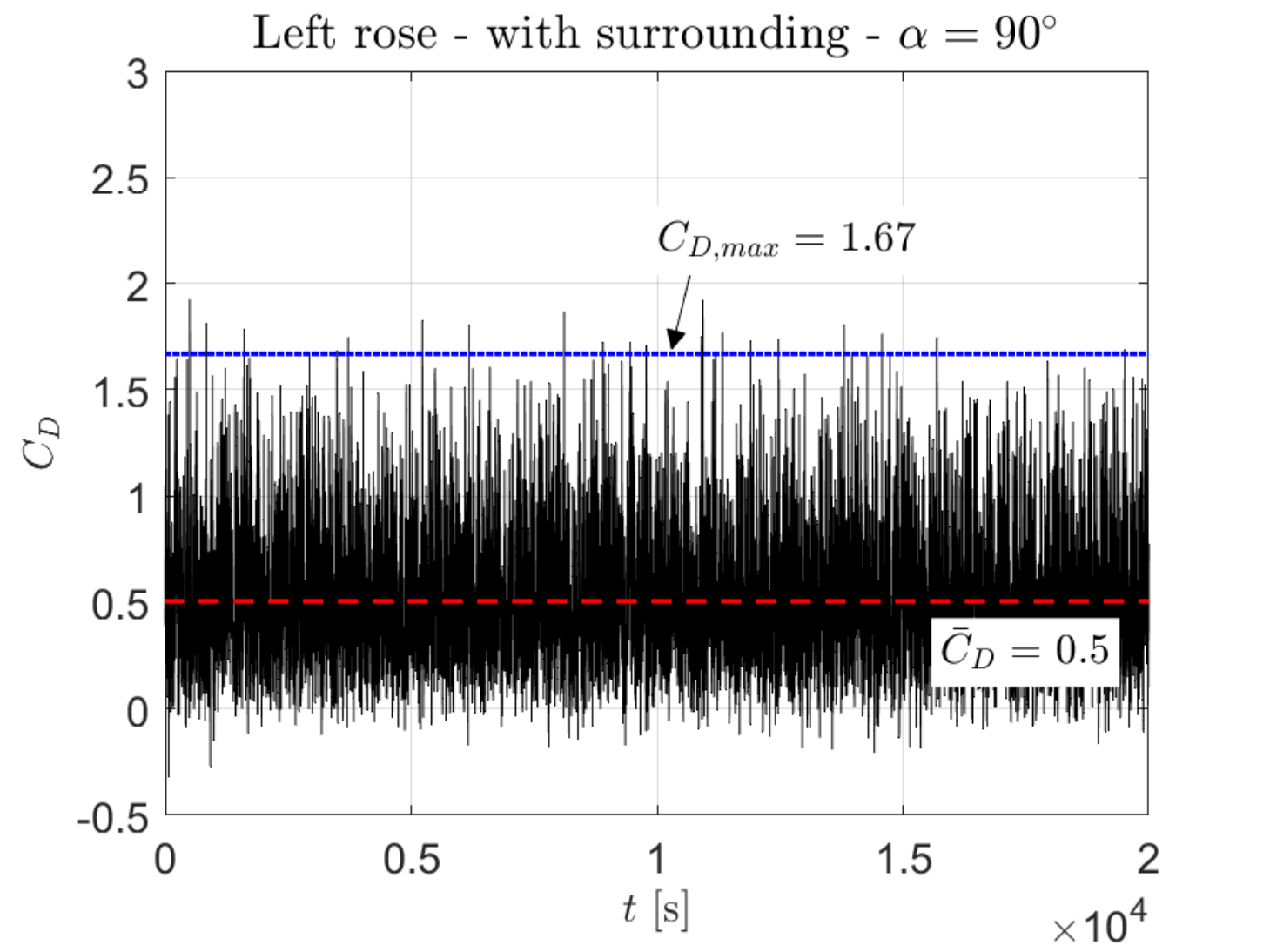}}
	\subfigure[\label{fig:Rosone_Surr_DX}]
  {\includegraphics[width=0.485\textwidth]{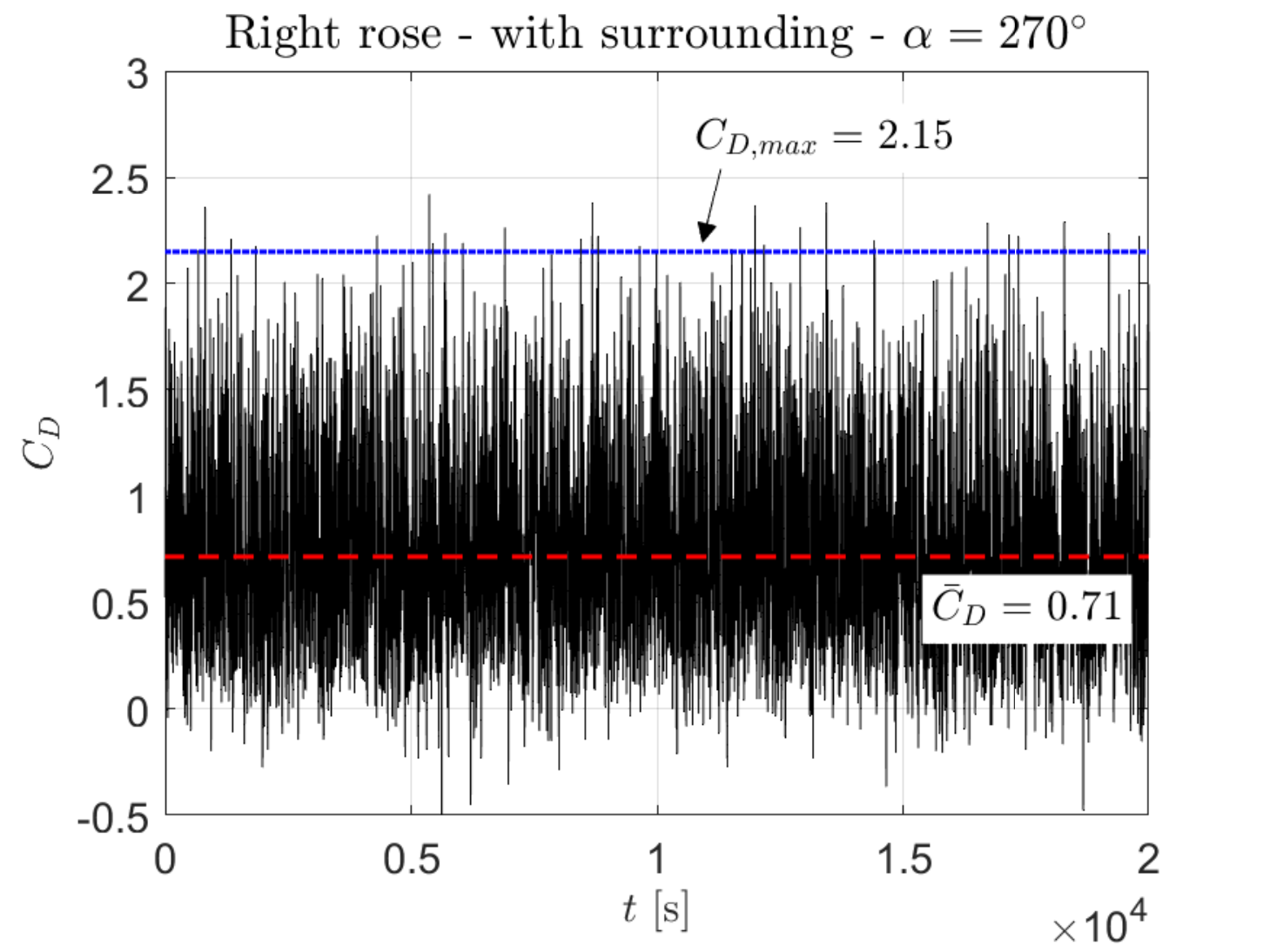}}
	\caption{Time history of the integrated force coefficient ($C_D = A^{-1}\int_A C_p dA$, where $A$ denotes the area of the window; time refers to full scale) on the great rose windows of the transept: (a) left rose, wind direction 90$^\circ$, without surrounding buildings; (b) left rose, wind direction 270$^\circ$, without surrounding buildings; (c) left rose, wind direction 90$^\circ$, with surrounding buildings; (d) right rose, wind direction 270$^\circ$, with surrounding buildings. Both mean and peak values of the load coefficient are reported.}
		\label{fig:rose_window}
	\end{center}
\end{figure}
%\end{comment}
Given the possible vulnerability of large Gothic rose window, as highlighted by Soissons's recent disaster mentioned in the Introduction, the pressures acting on the great roses of the transept have been integrated at each time instant to account for the non-perfect simultaneity of the fluctuations. The center of the rose windows is at a height above the ground of 25.8~m, and these have a surface area of about 85~m$^2$. A specific set of measurements has been used to this purpose, where the two facades of the transept are finely instrumented. Specifically, the pressure field on the rose windows is discretized with 30 pressure taps. The resulting time histories are shown in Fig.~\ref{fig:rose_window} for the wind directions perpendicular to the transept façade ($90^\circ$ and $270^\circ$).
It is apparent that the maximum load is obtained in the case without surrounding for the windward window. The load reduces by about 40\% when the surrounding portion of the city is modeled and the wind blows from North-Northeast (90$^\circ$-wind direction), due to the sheltering effect of the buildings very close to the Cathedral. Nevertheless, the load reduction is only 13\% when the wind comes from South-Southwest (270$^\circ$-wind direction) due to larger distance of the surrounding buildings. The leeward window is less loaded, and the resulting mean suction in the presence of the surrounding model is nearly the same as without surrounding for the 270$^\circ$-wind direction ($C_D = -0.43$), while a remarkable reduction is observed for the 90$^\circ$-wind direction ($C_D = -0.34$).
On the other hand, load fluctuation increases in the presence of the surrounding model and the peak force is even the same as in the case of a generic urban wind profile for a wind direction of 270$^\circ$.

\section{Discussion}
\subsection{General considerations}

\begin{figure}[t]
	\begin{center}
  \includegraphics[width=0.6\textwidth]{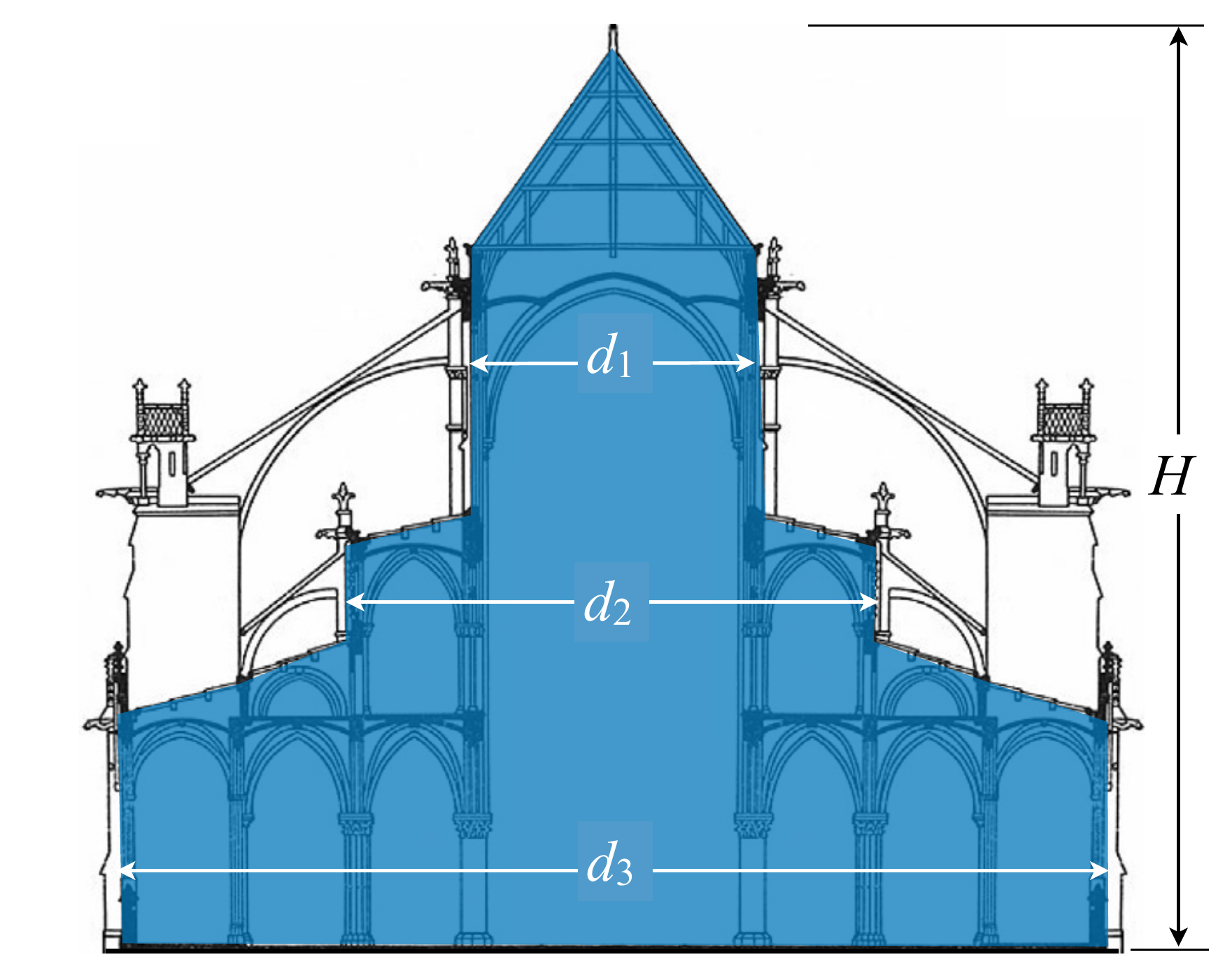}
	\caption{Schematic of a transversal cross section of the Cathedral.}
		\label{fig:Schema_sezione}
	\end{center}
\end{figure}

The results reported in the previous section revealed that the Cathedral's geometrical complexity reflects on the measured pressure field. Apart from the remarkable peculiarity of the apse and the main façade of the Cathedral, for the sake of simplification one may say that the Cathedral of Notre Dame is a long construction with a depth discontinuously varying with the height (Fig.~\ref{fig:Schema_sezione}), so that the height-to-depth ratio is small (close to 1) considering the lower part of the church but significantly higher (about 2.3) referring to the top of the vertical walls. Also, the roof presents a high pitch ($\sim55^\circ$), and there are several other architectural elements (such as the jutting out body of the transept, the large flying buttresses, the balustrade at the base of the roof, or the various orders of gables) that are supposed to significantly influence the aerodynamic behavior of the construction. It is clear that the urban integration of the structure also plays a key role.

From the practical engineering standpoint, however, it is important to understand if all the specific aerodynamic features of a large Gothic Cathedral lead to significantly different wind loads for design or safety verification purposes compared to a more-or-less ordinary building. It is also to be noticed that only canonical wind velocity directions have been analyzed so far (those parallel or perpendicular to the Cathedral's symmetry plane), while all possible wind directions must be considered instead for engineering purposes. Considering all these aspects, in the next section the load envelopes obtained for the finely-instrumented representative portion of the flank of the Cathedral (see Section~\ref{fascia_2D}) are compared with the few data available in the literature and with the loads proposed by Eurocode~1 for standard buildings.
%\begin{comment}
%\end{comment}

\subsection{Comparison with load models available in the literature}
\label{sec:comparison}

\begin{figure}[t]
	\begin{center}
	\subfigure[\label{fig:Comp_litt_no_surr}]
  {\includegraphics[width=0.485\textwidth]{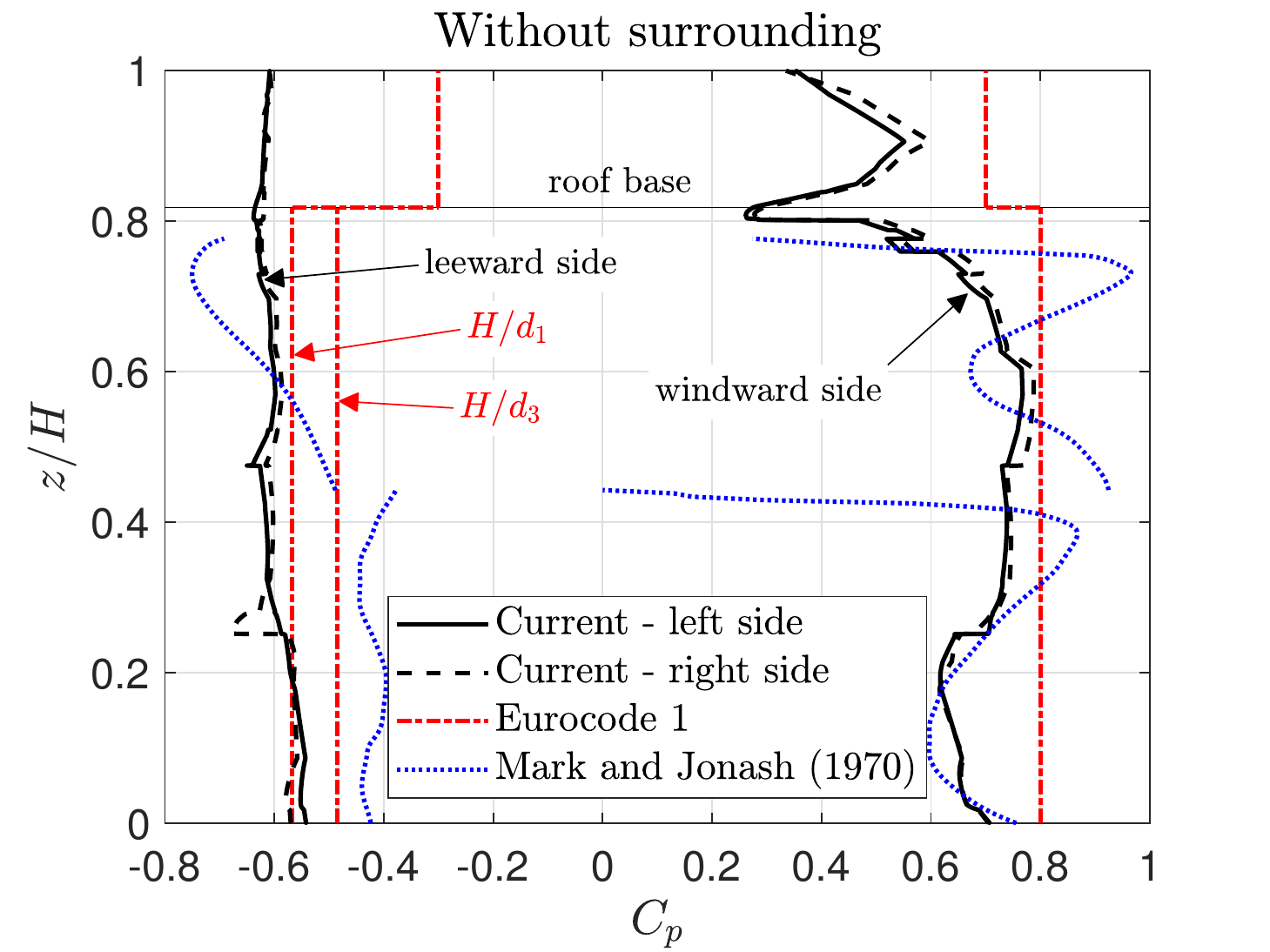}}
	\subfigure[\label{fig:Comp_litt_surr}]
  {\includegraphics[width=0.485\textwidth]{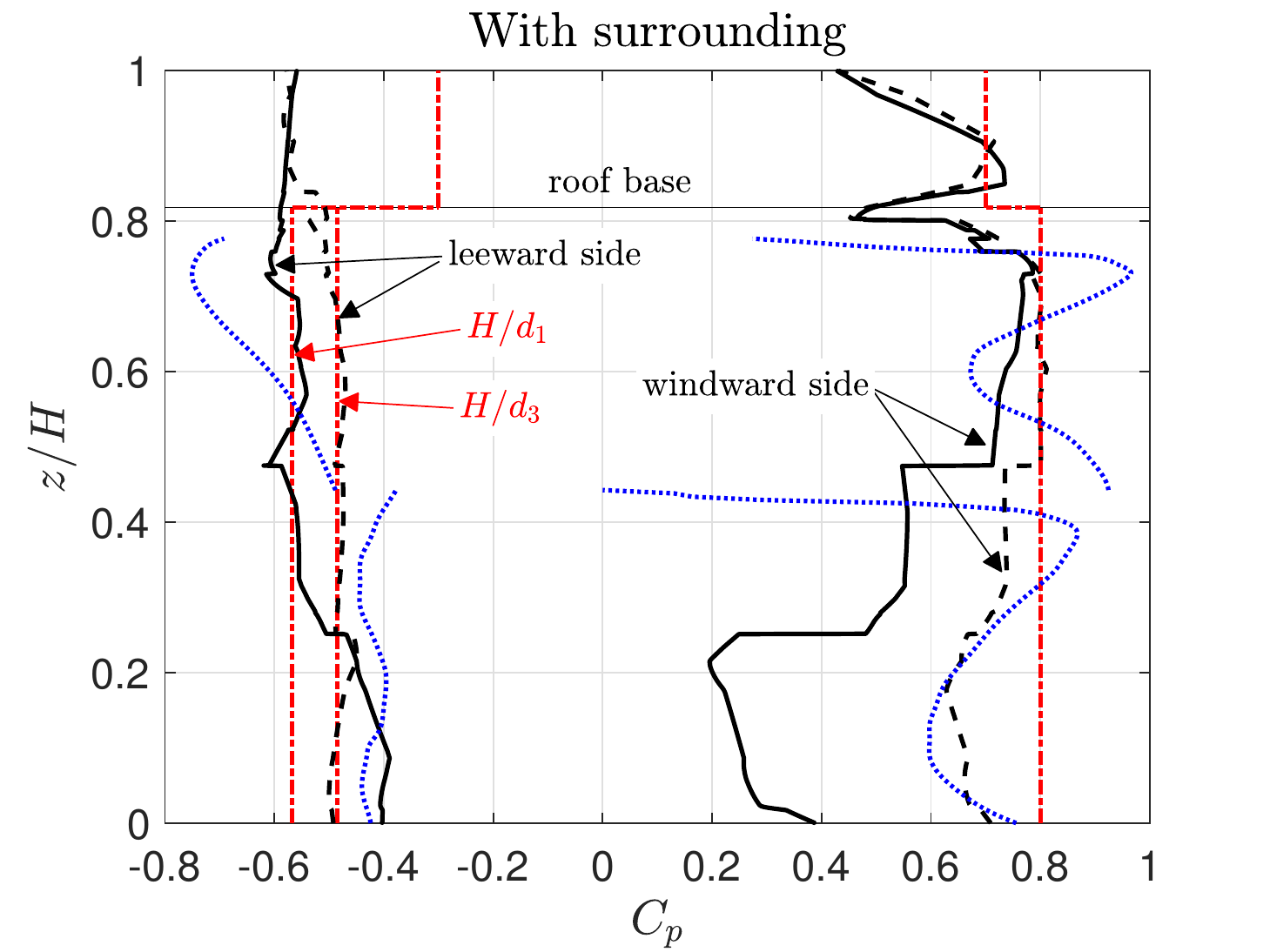}}
	\caption{Envelope mean pressure coefficient along a section of the Cathedral perpendicular to its longitudinal axis for the cases without (a) and with surrounding buildings (b). Comparison with Eurocode~1, \citet{Eurocode1}, prescriptions for ordinary buildings and with the data adapted from Mark and Jonash (1970), \citet{mark70}.}
	\label{fig:Comp_litt}
	\end{center}
\end{figure}

The mean pressure coefficients obtained for the finely instrumented portion of the flank of the Cathedral and the roof (Section~\ref{fascia_2D}) were averaged along the direction parallel to the longitudinal axis of the Cathedral (i.e., along the width of the considered Cathedral \textquotedblleft slice\textquotedblright). Moreover, to comply with the prescriptions of Eurocode~1, \citet{Eurocode1}, the envelope pressure coefficients were calculated at any height above the ground for wind azimuths between $-45^\circ$ and $+45^\circ$ around the direction perpendicular to one of the Cathedral's flanks. Results are reported in Fig.~\ref{fig:Comp_litt}. It is noteworthy that, in the case of a generic urban wind profile (without surrounding), the maximum mean pressure coefficient on the windward side is obtained for small or null skewness of the wind direction with respect to the canonical azimuths (either $90^\circ$ or $270^\circ$); therefore, the results for a wind perpendicular to the lateral walls represent a good estimate of the worst load case. In contrast, for the leeward side, the canonical wind directions provide nearly the lower load, whereas the higher suctions are obtained for inclinations of about $45^\circ$ with respect to them.
Clearly, results are more complicated and rather different for two flanks of the Cathedral when the surrounding buildings are modeled.
%\begin{comment}
%\end{comment}
The comparison of the results for the configurations with and without surrounding emphasizes the importance of modeling the buildings, the river and the other details of the city around the Cathedral.
As expected, for the case without surrounding, both windward and leeward pressure coefficient envelopes on the right and left sides of the Cathedral are nearly identical.
On the windward side, in the presence of the surrounding model, the pressure coefficient envelope exhibits significantly lower values in the inferior part of the walls when one considers the left flank of the Cathedral (then, when the wind direction is between $45^\circ$ and $135^\circ$, and the buildings are close to the church; see Fig.~\ref{fig:5}). In contrast, on the right flank (then, if the wind blows from directions $225^\circ$ and $315^\circ$), the pressure coefficients are very similar to the case without surrounding in the lower part of the walls. Above about 60\% of the roof top height, the pressure envelopes for the left and right flanks of the Cathedral become nearly identical and non-negligibly higher than the case without surrounding.
As for the leeward pressure coefficient envelopes, the suction is lower if the surrounding buildings are modeled, though there is a non-negligible difference between the left and the right side of the Cathedral.

The measured pressure coefficients on the windward side of the considered part of the Cathedral are generally lower (or even much lower in some regions) than the values suggested by Eurocode~1, except for a small portion of the roof for the case with surrounding. In contrast, on the leeward side the measured pressures tend to be slightly lower (higher suction) on the vertical walls of the Cathedral and significantly lower on the roof, especially for the configuration without surrounding.
Finally, the experimental data by Mark and Jonash (1970) for a Cathedral-like construction, \citet{mark70} are in agreement with the pressure coefficient envelopes in the lower part of the windward vertical walls for the case of a generic urban boundary layer, while they overestimate the loads in the upper part. In contrast, on the leeward side these literature data underestimate the suction on most part of the vertical walls, but they overestimate it in the top part.

Considering again the large roses of the transept (see Fig.~\ref{fig:rose_window}), one can remark that the mean load without surrounding on the windward window is in line with the prescriptions of Eurocode~1 (Fig.~\ref{fig:Schema_sezione}), provided that the velocity pressure at the top of the roof is used to normalize the force or the pressure coefficient. The gust factor (ratio of peak to mean load) for the windward rose perfectly complies with the value suggested by Eurocode~1 based on the turbulence intensity in the approaching wind profile at the height of the window (slightly less than 2.2). In contrast, it is higher in the presence of the surrounding buildings (larger than 3). However, it is very important to stress that the canonical wind directions (either $90^\circ$ or $270^\circ$) considered in the analysis of the load on the rose windows (see Fig.~\ref{fig:rose_window}) do not always represent the worst case scenario, especially for negative pressures on the leeward side, as it has already been emphasized for the finely instrumented portion of the flank of the Cathedral. In particular, mean suction forces as low as $-0.63$ (for a wind direction of either $135^\circ$ or $225^\circ$) and $-0.73$ (for a wind direction of $130^\circ$) have been obtained for the cases without and with surrounding, respectively.

Finally, one can notice that the effective load acting on most parts of the Cathedral will also depend on the internal pressures, which have not been investigated in the present work. Their estimate is a very complicated task as it requires the accurate modeling of diffused openings all over the building. However, in the absence of dominant openings (produced, for instance, by the failure of a window), for structural verifications it may be reasonable to consider the standard range $-0.3$ to $+0.2$ recommended by Eurocode~1, \citet{Eurocode1}.

\section{Conclusions}
\label{sec:conclusions}

The results of the wide wind tunnel test campaign that has been carried out on a scale model of the Cathedral of Notre Dame in Paris led to some major conclusions, hereafter reported.

Firstly, the complex geometry of a large Gothic cathedral, both in the aerial projection and in elevation, as well as the important aerodynamic role played by some architectural elements imply some specific phenomena in the wind-structure interaction process that do not allow the correct estimation of the wind loads based on the data available for standard buildings. Specifically, the present results reveal that, depending on the structural element considered or the specific configuration examined, these loads can be either significantly higher or lower than the values that can be predicted based on codes and standards.

Secondly, the influence on the wind load of the neighboring portion of the city where the considered Cathedral stands is extremely important, and can vary on a case-by-case basis. The current analysis highlights that the presence of the surrounding buildings does not only affect the aerodynamics of the portions of the Cathedral at a comparable height but also of parts significantly above (e.g., the roof).

Finally, this work may represent a first step towards the goal of guaranteeing the wind safety of architectural treasures like Gothic Cathedrals, even if in the future heavier and heavier effects of climate changes are likely to occur.

\bigskip

\section*{ Acknowledgements}
Authors want to thank:
\begin{itemize}
\item Martin Peters, former director of the Eiffel Wind Tunnel Laboratory, Paris, for his valuable help   concerning the history of wind tunnel research on historical monuments.
\item Emmanuel Portet and Laure Frèrejean, University of Versailles and Saint Quentin, for their fruitful commitment and help in the administrative and financial aspects of this research.
\item Bernardo Nicese, Clara Gessl and Petar Melnjak, who helped us in the execution of the laboratory tests in Prato.
\end{itemize}

%%%%%%%%%%% BIBLIOGRAPHY %%%%%%%%%%%%%%
\bibliographystyle{chicago} 
\bibliography{bibliowind}   % name your BibTeX data base

\end{document}